\documentclass[twocolumn,aps,prb,showpacs,amsmath,amssymb,floatfix]{revtex4}
\usepackage{graphics}
\usepackage{epsf,graphicx,pstricks,epsfig,psfrag}
\begin{document}
\newcommand{\comment}[1]{!!! \textbf{#1} !!!}
\title{Charge-density-wave formation in a half-filled 
fermion-boson transport model:\\ a  projective renormalization approach} 
%projector-based
%{Charge-density-wave formation in a fermion-boson model}
\author{S.~Sykora$^{1}$, K.~W.~Becker$^{2}$ and H. Fehske$^{3}$}
\affiliation{$^{1}$Center for Materials Theory,
Rutgers University, Piscataway, New Jersey 08855, USA\\
$^{2}$Institut f\"{u}r Theoretische Physik,
 Technische Universit\"{a}t Dresden, D-01062 Dresden, Germany\\
$^{3}$Institut f{\"ur} Physik,
             Ernst-Moritz-Arndt-Universit{\"a}t Greifswald,
             D-17489 Greifswald,
             Germany}
% \affiliation{
%   Institut f\"{u}r Theoretische Physik,
%  Technische Universit\"{a}t Dresden, D-01062 Dresden, Germany} 
% \author{H. Fehske}
% \affiliation{Institut f{\"ur} Physik,
%              Ernst-Moritz-Arndt-Universit{\"a}t Greifswald,
%              D-17489 Greifswald,
%              Germany}
{\today}


%\maketitle
%%%%%%%%%%%%%%%%%%%%%%%%%%%%%%%%%%%%%%%%%%%%%%%%%%%%%%%%%%%%%%%%%%%%%%%%%%%%%%%

\begin{abstract}
We study the metal-insulator transition in a very general two-channel 
transport model, where charge carriers are coupled to a correlated
background medium. The fluctuations of the background were described
as bosonic excitations, having the ability to relax. Employing an analytical 
projector-based renormalization technique, we calculate the ground-state
and spectral properties of this fermion-boson model and corroborate recent 
numerical results, which indicate---in dependence on the `stiffness' 
of the background medium---a Luttinger-liquid to charge-density-wave 
transition for the one-dimensional half-filled band case.  
%[Ejima et. al. Phys. Rev. Lett. {\bf 102}, 106404 (2009)].
In particular, we determine the renormalized electron and boson 
dispersion relations and show that the quantum phase transition is not 
triggered by a softening of the boson modes. Thus the charge density wave 
is different in nature from an usual Peierls distorted state.   
\end{abstract}
\maketitle

%%%%%%%%%%%%%%%%%%%%%%%%%%%%%%%%%%%%%%%%%%%%%%%%%%%%%%%%%%%%%%%
\section{Introduction}
Charge density waves (CDWs) are broken symmetry states of metals that 
predominantly appear in materials which have a highly anisotropic 
crystal and electronic structure~\cite{Gr94}. The formation of CDWs 
strongly depends on the band-filling and on the topology of the Fermi 
surface. Concerning the latter, one-dimensional (1D) systems are 
peculiar, because their Fermi surface consists of two points only. 
Not surprisingly, electron-electron and electron-phonon interactions, 
which are the driving forces behind most metal-insulator transitions, 
have more impact in reduced dimensions. At the same time, however, 
quantum fluctuations and finite-temperature effects---both 
counteracting any development of long-range order---become increasingly  
important as well. Then, all in all, in 1D, usually a rather complex 
interplay between the charge, spin, orbital and lattice degrees of freedom 
evolves, which, of course, will strongly affect the transport 
properties of the system. Prominent examples are quasi-1D halogen-bridged 
transition metal complexes (MX-chains)~\cite{BS93}.

The theoretical description of such highly correlated 1D systems 
is frequently based on `microscopic' models  like 
the half-filled SSH~\cite{SSH79}, Holstein~\cite{Ho59a,BMH98,HWBAF06}, 
Peierls~\cite{Pe55}, Hubbard~\cite{Hu63}, (quarter-filled) 
$t$-$J$ models~\cite{MSU98}, or combination of 
these~\cite{TC03}. Thereby, the complexity 
of the electron-electron or electron-phonon interactions 
in the Hamiltonians usually prevents the 
exact (numerical) solution of the model in the thermodynamic limit, 
which would be necessary in order to pinpoint a true quantum 
phase transition between a metal and a CDW. 

To some extent, the state of affairs improves if one considers 
simplified transport models instead, where particle motion takes 
place in an effective background medium. The `background'
reflects the correlations inherent in the system, e.g., 
the charge-, orbital- or spin-order in a solid. That quasiparticles
move through an ordered insulator is a very general situation in 
condensed matter physics~\cite{WOH09,Be09}. This scenario also applies
to soft matter systems like DNA, where the 
charge transport on the backbone is affected by the `configuration' of 
chemical side groups, which, vice versa, depends on the 
physical presence of charge carriers~\cite{KKB02}. 

Along this line, a novel fermion-boson quantum transport model, 
\begin{equation}\label{hem}
H =  -t_b^{} \sum_{\langle i, j \rangle}  c_j^{\dagger} c_i^{} (b_i^{\dagger}
+ b^{}_j) - \lambda_b^{} \sum_i (b_i^{\dagger} + b_i^{}) 
+ \omega_b^{} \sum_i b_i^{\dagger} b_i^{}\,, 
\end{equation}
has been proposed a few years ago~\cite{Ed06}, and was shortly afterwards 
solved for a single particle ($N_e=1$) in a 1D infinite system~\cite{AEF07}
by a variational numerical diagonalization technique~\cite{AEF07}.  
The Hamiltonian~(\ref{hem}) mimics the `background' by bosonic degrees
of freedom [$b_i^{(\dagger)}$], which influence and even may control the 
transport of fermionic particles [$c_i^{(\dagger)}$] on the 
$N$ sites of a regular lattice.  Every time a particle hops 
between nearest-neighbor 
Wannier sites $\langle i,j\rangle$, it creates (or destroys) a 
local excitation of energy $\omega_b$ in the background medium at the 
site it leaves (it enters). Clearly these distortions tie the 
particle to its origin (cf. the string effect as a hole moves
in a N\'{e}el spin background). Of course, any distortion of 
the background can heal out by quantum fluctuations (again, 
one can have spin fluctuations in mind). Accordingly, 
in the Hamiltonian~(\ref{hem}), the $\lambda_b$-term was included,   
allowing for spontaneous boson creation and annihilation processes.

Strong correlations, nevertheless, may evolve in a system 
described by the Hamiltonian~(\ref{hem}), provided the background 
excitations have a rather large energy and the ability of the background 
medium to relax is small, i.e., 
\begin{equation}
\Omega=\frac{\omega_b}{t_b} \gg 1 \quad\mbox{and}\quad
\Lambda=\frac{\lambda_b}{t_b} \ll 1\,. 
\end{equation}
For the half-filled band sector ($N_e=N/2$), these
correlations may even drive the system into an insulating 
state by establishing CDW long-range order. This has been
shown quite recently for the 1D case: small cluster 
diagonalizations~\cite{WFAE08} and density matrix renormalization group
(DMRG) calculations supplemented by finite-size scaling~\cite{EHF09,EF09b}
give strong evidence for a Tomonaga-Luttinger-liquid (TLL) 
CDW quantum phase transition
as $\Lambda$ becomes small at large enough $\Omega$. These purely numerical 
approaches rely on an (inevitable) truncation of the bosonic Hilbert space. 
Determining the metal-insulator phase boundary this seems to be uncritical,
because the CDW found at half-filling is a few-boson state. The situation
becomes more difficult if we enter the fluctuation-dominated regime of 
small $\Omega$, where many bosons are excited in the system.

In the present work, we investigate the fermion-boson transport 
model~\eqref{hem} by means of an analytical approach, which avoids these 
disadvantages. This approach, called projective renormalization method 
(PRM)~\cite{HSB08}, is based on a sequence of discrete unitary 
transformations, so that---in contrast to continuous (e.g. flow-equation based)
unitary transformation schemes~\cite{We94}---a direct link 
to perturbation theory 
can be provided. The method has already been successively 
applied to a number of many-particle 
models~\cite{HB05,SHBWF05,BSZ07,HSB08}. 
Here we will analyze the ground-state and spectral properties 
of the Hamiltonian~\eqref{hem} exclusively for the half-filled band case, in both the metallic and insulating
regimes. In particular we study the signatures of the 
TLL-CDW transition in terms of the renormalized quasiparticle
band and boson dispersion, and the boson spectral function. The paper is 
organized as follows. In Sec.~II.~A we briefly resume the basic  
concepts behind the PRM approach. The application of the PRM to the 
fermion-boson transport model will be described in detail in Sec.~II~B.  
Section III presents the results of the numerical evaluation of the
renormalization equations. We conclude in Sec.~VI.

%%%%%%%%%%%%%%%%%%%%%%%%%%%%%%%%%%%%%%%%%%%%%%%%%%%%%%%%%%%%%%%%%%%%%%%%%%%%%%%
%\section{Projector-based renormalization method (PRM)}
%%%%%%%%%%%%%%%%%%%%%%%%%%%%%%%%%%%%%%%%%%%%%%%%%%%%%%%%%%%%%%%%%%%

%\textit
\section{Theoretical approach}
\subsection{Projector-based renormalization method}
The PRM starts from the usual decomposition of a many-particle 
Hamiltonian into a solvable unperturbed part $H_0$ and a 
perturbation $H_1$, where $H_1$ should not contain any part 
that commutes with $H_0$. Thus, the perturbation $H_{1}$ 
consists of transitions between the eigenstates of $H_0$ with 
non-vanishing transition energies. The basic idea of the PRM is to construct an effective 
Hamiltonian $H_\lambda = H_{0,\lambda} + H_{1,\lambda}$
with renormalized parts $H_{0,\lambda}$ and $H_{1,\lambda}$,
where all transitions with energies 
$|E_{0,\lambda}^n -E_{0,\lambda}^m|$ larger than a given cutoff energy 
$\lambda$ are eliminated. $E_{0,\lambda}^n$ and $E_{0,\lambda}^m$ 
denote the eigenenergies of $H_{0,\lambda}$. 

The renormalization procedure starts from the cutoff energy $\lambda=\bar{\lambda}$ 
of the original model $H$ and proceeds in steps of $\Delta \lambda$ 
to lower values of $\lambda$. Every renormalization step is performed by 
means of a unitary transformation,
%\comment{'The elimination 
%between the cutoffs $\lambda$ and $(\lambda - \Delta \lambda)$
%reads ' wuerde ich weglassen}
\begin{eqnarray}
  \label{2}
  H_{\lambda-\Delta\lambda} = 
  e^{X_{\lambda,\Delta\lambda}} \, H_{\lambda} \,
  e^{-X_{\lambda,\Delta\lambda}}.
\end{eqnarray}
The generator $X_{\lambda,\Delta \lambda}$  of the unitary
transformation has to be fixed appropriately (for details see 
Ref.~\onlinecite{HSB08}). For instance,  in lowest order perturbation theory, 
it reads 
\begin{eqnarray}
  \label{3}
 X_{\lambda,\Delta\lambda} &=& \frac{1}{{\bf L}_{0, \lambda}} 
 {\bf Q}_{\lambda - \Delta \lambda}{H}_{1, \lambda}\, .
\end{eqnarray}
Here, ${\bf L}_{0, \lambda}$ is the Liouville superoperator of the 
`unperturbed' 
Hamiltonian ${H}_{0, \lambda}$, which is defined by the commutator
of ${H}_{0, \lambda}$ with any operator variable $A$,
i.e.
${\bf L}_{0, \lambda} A = [H_{0, \lambda}, A]$, 
and $ {\bf Q}_{\lambda - \Delta \lambda}$ is a projection superoperator, 
which projects on all transitions with respect to the eigenspectrum of 
${H}_{0, \lambda -\Delta \lambda}$ with 
transition energies larger than $\lambda - \Delta \lambda$.  
 In this way difference equations can be derived which 
connect the parameters of $H_\lambda$ with those of 
$H_{\lambda - \Delta \lambda}$, and which are called renormalization 
equations.

The limit $\lambda \rightarrow 0$ provides the desired effective Hamiltonian 
$ 
  \tilde{H}= H_{\lambda \rightarrow 0} 
  = H_{0, \lambda \rightarrow 0}
$
where the elimination of the transitions originating from the perturbation 
$H_{1}$ leads to a renormalization of the parameters of 
$\tilde{H}$. Note that $\tilde{H}$ 
is diagonal or at least quasi-diagonal and allows to evaluate
physical quantities. The final results depend on the parameter values of the 
original Hamiltonian $H$. Finally, we note that $\tilde{H}$ and 
$H $ have the same eigenvalue problem since both Hamiltonians are 
connected by a unitary transformation.  

To evaluate expectation values of operators $A$, 
formed with the full Hamiltonian,
% for operator variables $A$, 
we have to apply the unitary transformation as well,
\begin{eqnarray}
\label{4}
 \langle A \rangle =
\frac{{\rm Tr} A e^{-\beta H}}{ {\rm Tr} e^{-\beta H}} = 
\langle A(\lambda) \rangle_{H_\lambda}=
 \langle \tilde{A}\rangle_{\tilde{H}} \, , 
\end{eqnarray}
where we define 
$A(\lambda) = e^{X_\lambda}Ae^{-X_\lambda}$ and $\tilde{A}= 
A(\lambda \rightarrow 0)$. Thus additional renormalization equations 
are required for  $A(\lambda)$. 

%%%%%%%%%%%%%%%%%%%ררר
%\textit
\subsection{Application to the two-channel transport model}
\subsubsection{Renormalization equations}
We first rewrite the model~\eqref{hem}, performing a unitary transformation
$b_i \mapsto b_i + \lambda_b / \omega_b$  that eliminates the boson relaxation 
term in favor of a free-particle hopping channel,    
\begin{equation}
  \label{1}
  H =
  - t_f \sum_{\langle i,j\rangle} c_{j}^\dagger c_{i}^{} 
- t_b \sum_{\langle i,j\rangle} \; 
    c_j^\dagger c_i^{}(b_i^\dagger + b_j^{})+ \omega_b \sum_i  \; b_i^\dagger b_i^{}\,,
\end{equation}
with 
\begin{equation}
\label{t_f_L}
\frac{t_f}{t_b}=2\frac{\lambda_b}{\omega_b}= 2\frac{\Lambda}{\Omega}\,.
\end{equation}
This makes the two transport channels 
contained in the Hamiltonian~\eqref{hem} explicit: the coherent 
particle transfer, which takes place on an energy 
scale $(\propto t_f)$, and the 
boson-affected hopping  $(\propto t_b)$. 

Next, in order to exploit the translation invariance, we consider 
the Hamiltonian~\eqref{1} in momentum space 
\begin{eqnarray}
\label{5}
H &=&   \sum_{k} \varepsilon_k c_{k}^{\dag} c_k^{} +
\omega_{b} \sum_{q}  b_{q}^{\dag} b_q^{}\nonumber\\  
&&+ \frac{1}{\sqrt{N}} \sum_{k,q} g_k
\big( b_{q}^{\dag} c_{k}^{\dag}
 c_{k + q}^{} + b_q^{} c_{k + q}^{\dag}
 c_{k}^{} \big)\,.
\end{eqnarray}
In what follows, we consider a 1D lattice with lattice constant $a$,
i.e., $\varepsilon_{k} = -2t_f \cos{ka}$ and $g_{k}= -2t_b \cos{ka}$. 

Going forward, it turns out to be useful to remove the mean-field part from 
the fermion-boson coupling term. Defining fluctuation operators,
\begin{eqnarray} 
\delta(c_k ^\dagger c_{k+q}^{}) = c_k ^\dagger c_{k+q}^{} - 
\langle c_k ^\dagger c_{k}^{}\rangle\, \delta_{q,0}\,,
\end{eqnarray}
the Hamiltonian \eqref{5} takes the form 
\begin{eqnarray}
\label{6}
H &=&   \sum_k 
\varepsilon_k^{} c_{k}^{\dag} c_{k}^{} +
\omega_{b}^{} \sum_{q}  b_{q}^{\dag} b_{q}^{}\nonumber\\
&&+
\frac{1}{\sqrt{N}} \sum_{k} g_k^{}
\langle c_{k}^{\dagger} c_{k}^{} \rangle
( b_{0}^\dagger +  b_0^{} ) \nonumber \\
  &&+
\frac{1}{\sqrt{N}} \sum_{k,q} g_{k}^{}
\big[ b_{q}^{\dag} \delta(c_{k}^{\dag}
 c_{k + q}^{}) + b_q^{} \delta(c_{k + q}^{\dag}
 c_{k}^{}) \big]\,.
\end{eqnarray}
Obviously, the purely bosonic part of $H$, i.e.~the second and third 
term of Eq.\eqref{6},
can be diagonalized by a shift of the bosonic operators. 
Introducing new bosonic creation  operators, 
  \begin{eqnarray}
\label{7}
B_q^{\dagger} = b_q^{\dagger} + \frac{1}{\sqrt{N}} \sum_k \frac{g_k}{\omega_b}
\langle c_{k}^{\dagger} c_{k} \rangle \delta_{q,0} \,,
\end{eqnarray}  
the Hamiltonian \eqref{6} can 
be rewritten as
$H =  H_0  + H_1$ 
with 
\begin{eqnarray}
 H_0&=&  \sum_{k} 
 \Big(\varepsilon_{k} -   2 g_k \frac{1}{N}    
\sum_{k'} \frac{g_{k'}}{\omega_b}
\langle c_{k'}^{\dagger} c_{k'}^{} \rangle \Big) \,
 c_{k}^{\dag} c_{k}^{}\nonumber\\ 
 &&+ \omega_{b}^{} \sum_{q}  B_{q}^{\dag} B_{q}^{}  
+ \frac{1}{N \omega_b}
\Big(  \sum_k g_k
\langle c_{k}^{\dagger} c_{k}^{} \rangle
\Big)^2, \\
 %%%%%%%
 \label{8}
 H_1   &=&
 \frac{1}{\sqrt{N}} \sum_{k, q} g_{k}
\Big[ B_{q}^{\dag} \delta(c_{k}^{\dag} c_{k + q}^{}) 
+ B_{q} \delta(c_{k + q}^{\dag}  c_k^{})\Big].   
\end{eqnarray}
Following the ideas of the PRM approach, we make  the following {\it ansatz} 
for the renormalized Hamiltonian ${H}_\lambda$ (after all transitions with energies larger than $\lambda$ have been integrated out), 
$H_\lambda =  H_{0,\lambda}  + H_{1,\lambda}$, 
where 
\begin{eqnarray}
\label{9}
 H_{0,\lambda}&=&  \sum_{k} 
 \varepsilon_{k,\lambda}   c_{k}^{\dag} c_{{k}}^{}  
 + \sum_k \Delta_{k,\lambda}  c_{k}^{\dag} c_{k+Q}\nonumber\\  
 &&+\sum_{q} \omega_{q,\lambda}  B_{q}^{\dag} B_{q}^{}  
+ E_\lambda \,,
\end{eqnarray}
\begin{equation}
\label{9b}
 H_{1,\lambda}   =
 \frac{1}{\sqrt{N}} \sum_{k, q} g_{k} \Theta_{k,q}(\lambda)
\Big[ B_{q}^{\dag} \delta(c_{k}^{\dag} c_{k + q}^{}) 
+ B_{q}^{} \delta(c_{k + q}^{\dag}  c_k^{}) \Big]\,.
\end{equation}
The $\Theta$-function in  \eqref{9b}
$$\Theta_{k,q}(\lambda)= 
\Theta(\lambda-| \varepsilon_{k,\lambda}-\varepsilon_{k +q,\lambda} 
+ \omega_{q,\lambda}|)$$ 
guarantees that only transitions with excitation energies smaller than $\lambda$
remain in $ H_{1,\lambda}$.  In $ H_{0,\lambda}$, also a
symmetry breaking field, $\Delta_{k,\lambda}$, was introduced which couples together particle-hole excitations with wave vectors $k$ and $k+Q$, where $Q=\pi/a$. Note that the renormalization of 
$\Delta_{k,\lambda}$  may lead to a transition to a CDW ground state at half filling. 

By integrating out all transitions between the cutoff  
of the original model $\bar{\lambda}$
and $\lambda=0$, all parameters of the original model 
will become renormalized. To find their 
$\lambda$-dependence, we derive renormalization 
equations for the parameters $\varepsilon_{k,\lambda}$, $\Delta_{k,\lambda}$,  
$ \omega_{q,\lambda} $, and $ E_\lambda$. The coupling parameter $g_k$ 
is not renormalized when we restrict ourselves to lowest order perturbation theory in one renormalization step. 
The initial parameter values are determined by the original model   
($\lambda=\bar{\lambda}$):
\begin{eqnarray}
\label{10}
\varepsilon_{k, \bar{\lambda}} &=&
\varepsilon_{k} -   2 g_k \frac{1}{N}    \sum_{k'} \frac{g_{k'}}{\omega_b}
\langle c_{k'}^{\dagger} c_{k'}^{} \rangle\,,\\
E_{\bar{\lambda}} &=& \frac{1}{N \omega_b}
\Big(  \sum_k g_k
\langle c_{k}^{\dagger} c_{k} \rangle\Big)^2\,,\\
\Delta_{k, \bar{\lambda}} &=&0^+\,, \;\;\mbox{and}\;\;\;
\omega_{q, \bar{\lambda}} \,=\, \omega_b \,.
\end{eqnarray}

Let us assume that the symmetry 
breaking field $\sim \Delta_{k,\lambda}$  can be considered as small compared
to the hopping part in 
${H}_{0, \lambda}$. 
In this case, the dynamics of ${H}_{0, \lambda}$ 
is approximately governed by
\begin{eqnarray}
\label{11}
\,[H_{0, \lambda},  c_k^\dagger] &=& \varepsilon_{k,\lambda } c_k^\dagger \,,\\
\,[H_{0, \lambda}, B_q^\dagger] &=& \omega_{q, \lambda} B_q^\dagger \,.
\end{eqnarray}
Following [\onlinecite{HSB08}], the lowest order expression
generator $X_{\lambda, \Delta \lambda}$ is obtained as 
\begin{eqnarray}
\label{12}
X_{\lambda, \Delta \lambda} &=& \frac{1}{\sqrt{N}} \sum_{k,q}
\frac{g_k \Theta_{k,q}(\lambda, \Delta \lambda)}
{\varepsilon_{k,\lambda} - \varepsilon_{k+q,\lambda} +\omega_{q, \lambda}} \nonumber\\
&&\times \Big[
B_q^\dagger \delta(c_k^\dagger c_{k+q}) - B_q \delta (c_{k+q}^\dagger c_k)
\Big]\,. 
\end{eqnarray}
Here,
\begin{eqnarray}
&&\Theta_{k,q}(\lambda, \Delta \lambda)=
\Theta(\lambda-| \varepsilon_{k,\lambda}-\varepsilon_{k +q,\lambda} 
+ \omega_{q,\lambda}|)\quad   \\
&&\;\;\;\;\; \times 
\Theta(| \varepsilon_{k,\lambda- \Delta \lambda}-
\varepsilon_{k +q,\lambda- \Delta \lambda} 
+ \omega_{q,\lambda - \Delta \lambda }| - (\lambda - \Delta \lambda))\nonumber
\end{eqnarray}
is a product of two $\Theta$-functions which assure that only excitations between $\lambda$ and $\lambda -\Delta \lambda$
are eliminated by the unitary transformation \eqref{2}. From  Eq.\eqref{2}, the Hamiltonian ${H}_{\lambda- \Delta \lambda}$ 
is easily evaluated within second order perturbation theory,  
\begin{eqnarray}
\label{13}
H_{\lambda - \Delta \lambda} &=& H_{\lambda}
+ [X_{\lambda, \Delta \lambda}, H_{0, \lambda} ]\nonumber\\&&\quad\;+ \frac{1}{2} [X_{\lambda, \Delta \lambda},
 [X_{\lambda, \Delta \lambda}, H_{0, \lambda}  ]   ]\nonumber\\
&&\quad\;
+ [X_{\lambda, \Delta \lambda}, H_{1, \lambda}]\,.
\end{eqnarray}

An alternative expression for  ${H}_{\lambda- \Delta \lambda}$ is obtained  by replacing $\lambda$ in  Eqs.\eqref{9}, \eqref{9b} by the reduced cutoff $\lambda - \Delta \lambda$. Thus,   
${H}_{\lambda- \Delta \lambda} =  {H}_{0, \lambda- \Delta \lambda} 
+  {H}_{1, \lambda- \Delta \lambda}$
with 
\begin{eqnarray}
\label{14}
 H_{0,\lambda - \Delta \lambda}&=&  \sum_{k} 
 \varepsilon_{k,\lambda - \Delta \lambda}   c_{k}^{\dag} 
c_{{k}}^{} + \sum_k \Delta_{k,\lambda- \Delta \lambda}  c_{k}^{\dag} c_{k+Q}^{}  
\nonumber\\  
&& +
 \sum_{q} \omega_{q,\lambda- \Delta \lambda}  B_{q}^{\dag} B_{q}^{}  
+ E_{\lambda- \Delta \lambda}, \\
 %%%%%%%%% 
\label{14b}
 H_{1,\lambda - \Delta \lambda}   &=&
 \frac{1}{\sqrt{N}} \sum_{k, q} g_{k} 
\Theta_{k,q}(\lambda-  \Delta \lambda)\nonumber\\
&&\times\Big[ B_{q}^{\dag} \delta(c_{k}^{\dag} c_{k + q}) 
+ B_{q} \delta(c_{k + q}^{\dag}  c_k) \Big]\,.  
\end{eqnarray}
A comparison of expression~\eqref{13} with~\eqref{14}, \eqref{14b} leads to 
the renormalization 
equations which connect the parameter values of the Hamiltonian 
at cutoff $\lambda$ with those at cutoff $\lambda - \Delta \lambda$. 
We obtain
\begin{widetext}
\begin{eqnarray}
\label{15}
\varepsilon_{k,\lambda - \Delta \lambda} - \varepsilon_{k,\lambda} &=&
\frac{1}{N} \sum_q \left\{ (n_q^B + n_{k+q}^c)
\frac{g_k^2 \Theta_{k,q}(\lambda, \Delta \lambda)}
{ \varepsilon_{k,\lambda} - \varepsilon_{k+q,\lambda} +\omega_{q, \lambda}} \right.  \left. +  (n_q^B - n_{k-q}^c +1)
\frac{g_{k-q}^2 \Theta_{k-q,q}(\lambda, \Delta \lambda)}
{ \varepsilon_{k,\lambda} - \varepsilon_{k-q,\lambda} -\omega_{q, \lambda}}\right\}, \\ 
\label{omeg}
 \omega_{q, \lambda - \Delta \lambda}- \omega_{q, \lambda} &=&
\frac{1}{N} \sum_k (n_k^c - n_{k+q}^c)
\frac{g_k^2 \Theta_{k,q}(\lambda, \Delta \lambda)}
{ \varepsilon_{k,\lambda} - \varepsilon_{k+q,\lambda} +\omega_{q, \lambda}}\,,\\
\Delta_{k, \lambda - \Delta \lambda} - \Delta_{k, \lambda} &=&
\frac{1}{N} \sum_q \left\{
\frac{g_k^2 \Theta_{k,q}(\lambda) \Theta_{k+Q,q}(\lambda, \Delta \lambda)}
{\varepsilon_{k+Q,\lambda}- \varepsilon_{k+q+Q, \lambda}+ \omega_{q, \lambda} } 
 +
\frac{g_{k+q}^2 \Theta_{k+q,-q}(\lambda) \Theta_{k+q+Q,-q}(\lambda, \Delta \lambda)}
{\varepsilon_{k+q+Q,\lambda}- \varepsilon_{k+Q, \lambda}+ \omega_{q, \lambda} } \right. \nonumber \\\label{16}
&& \left. \hspace*{1.2cm} +
\frac{g_k^2 \Theta_{k+Q,q}(\lambda) \Theta_{k,q}(\lambda, \Delta \lambda)}
{\varepsilon_{k,\lambda}- \varepsilon_{k+q, \lambda}+ \omega_{q, \lambda} }
+ 
\frac{g_{k+q}^2 \Theta_{k+q+Q, -q}(\lambda) \Theta_{k+q,-q}(\lambda, \Delta \lambda)}
{\varepsilon_{k+q,\lambda}- \varepsilon_{k, \lambda}+ \omega_{q, \lambda} } \nonumber 
\right\} d_{k+q}^c \nonumber \\
&& 
- \frac{1}{N} \sum_{k'} \left\{
\frac{\Theta_{k+Q, Q}(\lambda) \Theta_{k',Q}(\lambda, \Delta \lambda)}
{\varepsilon_{k',\lambda}- \varepsilon_{k'+Q, \lambda}+ \omega_{Q, \lambda} } 
+
 \frac{\Theta_{k, Q}(\lambda) \Theta_{k'+ Q,Q}(\lambda, \Delta \lambda)}
{\varepsilon_{k'+ Q,\lambda}- \varepsilon_{k', \lambda}+ \omega_{Q, \lambda} }  
\nonumber \right. \\\label{16b}
&& \left. \hspace*{1.4cm} -
\frac{\Theta_{k', Q}(\lambda) \Theta_{k,Q}(\lambda, \Delta \lambda)}
{\varepsilon_{k,\lambda}- \varepsilon_{k+Q, \lambda}+ \omega_{Q, \lambda} }  
-
\frac{\Theta_{k'+Q, Q}(\lambda) \Theta_{k+Q,Q}(\lambda, \Delta \lambda)}
{\varepsilon_{k+Q,\lambda}- \varepsilon_{k, \lambda}+ \omega_{Q, \lambda} } 
\right\} g_k g_{k'} d_{k'}^c\,.
\end{eqnarray}
\end{widetext}
Since the equation for the energy shift 
$E_{\lambda - \Delta \lambda}$ is  not needed in the 
following it has been left out for briefness. 
In Eqs.\eqref{15}--\eqref{16}, we have defined new expectation values
$n_k^c = \langle c_k^\dagger c_{k}^{} \rangle$, 
$n_q^B = \langle B_q^\dagger B_{q^{}} \rangle$, and 
$d_k^c= \langle c_k^\dag c_{k+Q}^{}\rangle$, 
which are formed with the full Hamiltonian $H$. 
Suppose these expectation values are known, 
the renormalization between the cutoff $\bar{\lambda}$
of the original Hamiltonian $H$ and $\lambda= 0$ leads to the Hamiltonian
$\tilde{H}$,  
\begin{equation}
\label{17}
 \tilde{H} = \sum_k \tilde{\varepsilon}_k c_k^\dagger c_k^{}
 + \sum_k \tilde{\Delta}_k c_k^\dagger c_{k+ Q}^{}
+ \sum_q \tilde{\omega}_q B_q^\dagger B_q^{} + \tilde{E}\,, 
\end{equation}
where 
$\tilde{\varepsilon}_k$, $\tilde{\Delta}_k$, 
$\tilde{\omega}_q$ and $\tilde{E}$ denote the
parameter values at $\lambda=0$.  

Note that the fully renormalized Hamiltonian $\tilde{H}$
describes an uncoupled system of renormalized (dressed) electrons and bosons. 
Both parts are 
quadratic either in the fermionic or in bosonic operators. By a
rotation in the fermionic subspace the electronic part of $\tilde{H}$ can
easily be diagonalized. Thus,  any expectation value 
can be evaluated. This property will be used in the following in order
to evaluate the yet unknown expectation values $n_k^c$, $n_q^B$, and $d_k^c$.

\subsubsection{Expectation values}

The expectation values can be evaluated self-consistently 
within the PRM formalism 
by applying the same unitary transformation as was used before 
for the Hamiltonian.  Following Eq.\eqref{2}, for instance, $n_k^c$ can be expressed by $n_k^c= \langle c_k^\dagger (\lambda) c_k(\lambda)^{} 
\rangle_{{H}_\lambda}$, where 
$\langle \cdots \rangle_{{H}_\lambda}$  means the average formed with 
${H}_\lambda$ and $c_k^\dagger(\lambda)$ is given by $c_k^\dagger(\lambda)= 
e^{X_\lambda} c_k^\dagger e^{-X_\lambda}$. For the transformed 
operators $c_k^\dagger(\lambda)$ and $B_q^\dagger(\lambda)$ 
we use the ansatz  
\begin{equation}
\label{18}
c_k^\dagger(\lambda) = \alpha_{k, \lambda} c_k^\dagger + 
\sum_q \left(\beta_{k,q, \lambda}^{} B_q^{} c_{k+q}^\dagger + 
\gamma_{k,q,\lambda}^{}B_q^\dagger 
c_{k-q}^\dagger \right)
\end{equation}
and
\begin{equation}
B_q^\dagger(\lambda) = \phi_{q,\lambda}^{} B_q^\dagger + \eta_{q,\lambda}^{} 
B_{-q} +\sum_k \psi_{k,q,\lambda}^{} \delta(c_{k+q}^\dagger c_k^{})\,,
\end{equation}
respectively.
The operator structure is again taken over from the lowest order expansion of the unitary transformation, apart from the second term in $B_q^\dagger(\lambda)$ 
which is due to higher-order terms. For the $\lambda$-dependent coefficients $\alpha_{k,\lambda},
\beta_{k,q,\lambda}, \cdots$ also renormalization equations have to be derived,
\begin{widetext}
\begin{eqnarray}
\label{alpha}
\alpha_{k, \lambda - \Delta \lambda} - \alpha_{k, \lambda} &=&
-\frac{1}{2N} \sum_q \left(n_{k+q}^c + n_q^B\right) \left(
\frac{g_k}{\varepsilon_{k, \lambda}- \varepsilon_{k+q, \lambda} + \omega_{q,\lambda}} \right)^2 \,
\alpha_{k, \lambda} \Theta_{k,q}(\lambda, \Delta \lambda) \nonumber \\
&&\quad- 
\frac{1}{2N} \sum_q \left(1- n_{k-q}^c + n_q^B\right) \left(
\frac{g_{k-q}}{\varepsilon_{k-q, \lambda}- \varepsilon_{k, \lambda} + \omega_{q,\lambda}} \right)^2 \,
\alpha_{k, \lambda} \Theta_{k-q,q}(\lambda, \Delta \lambda)\,,\\
\beta_{k,q, \lambda - \Delta \lambda} - \beta_{k,q, \lambda} &=&
-\frac{1}{\sqrt N} 
\frac{g_k}{\varepsilon_{k, \lambda}- \varepsilon_{k+q, \lambda} + \omega_{q,\lambda}}  \,
\alpha_{k, \lambda} \Theta_{k,q}(\lambda, \Delta \lambda)\,, \\
\gamma_{k,q, \lambda - \Delta \lambda} - \gamma_{k,q, \lambda} &=&
\frac{1}{\sqrt N} 
\frac{g_{k+q}}{\varepsilon_{k+q, \lambda}- \varepsilon_{k, \lambda} + \omega_{q,\lambda}}  \,
\alpha_{k, \lambda} \Theta_{k+q,-q}(\lambda, \Delta \lambda)\,, \\
\label{phi}
\phi_{q, \lambda - \Delta \lambda} - \phi_{q, \lambda} &=&
-\frac{1}{2N} \sum_k \left(n_{k}^c - n_{k+q}^c\right) \left(
\frac{g_k}{\varepsilon_{k, \lambda}- \varepsilon_{k+q, \lambda} + \omega_{q,\lambda}} \right)^2 \,
\phi_{q, \lambda} \Theta_{k,q}(\lambda, \Delta \lambda)\,,\\
\label{eta}
\eta_{q, \lambda - \Delta \lambda} - \eta_{q, \lambda} &=&
\frac{1}{2N} \sum_k \left(n_{k}^c - n_{k+q}^c\right) \left(
\frac{g_{k+q}}{\varepsilon_{k+q, \lambda}- \varepsilon_{k, \lambda} 
+ \omega_{q,\lambda}} \right)^2 \,
\eta_{q, \lambda} \Theta_{k+q,-q}(\lambda, \Delta \lambda)\,, \\
\label{psi}
\psi_{k,q, \lambda - \Delta \lambda} - \psi_{k,q, \lambda} &=&\label{19}
-\frac{1}{\sqrt N}\left[ 
\frac{g_k}{\varepsilon_{k, \lambda}\!-\! \varepsilon_{k+q, \lambda} \!+\! \omega_{q,\lambda}}  \,
\phi_{q, \lambda} \Theta_{k,q}(\lambda, \Delta \lambda) -
%\frac{1}{\sqrt N} 
\frac{g_{k+q}}{\varepsilon_{k+q, \lambda}\!-\! \varepsilon_{k, \lambda} \!+\! \omega_{q,\lambda}}  \,
\eta_{q, \lambda} \Theta_{k+q,-q}(\lambda, \Delta \lambda)\right] .
\end{eqnarray}
\end{widetext}
Integrating these equations between $\bar{\lambda}$ (where 
$\alpha_{k,\bar{\lambda}}=
\phi_{q,\bar{\lambda}}=1$ and all other coefficients zero)
 and $\lambda=0$, we arrive at the final result for $n_k^c$, $n_q^B$, 
and $d_k^c$:   
 \begin{eqnarray}
\label{20a}
n_k^c \!&=&\! \tilde{\alpha}_k^2 \tilde{n}_k^c \\
&&+  \sum_q \Big[ \tilde{\beta}_{k,q}^2 (1+\tilde{n}_q^B)
\tilde{n}_{k+q}^c + \tilde{\gamma}_{k,q}^2 \tilde{n}_q^B \tilde{n}_{k-q}^c 
\Big],\nonumber \\
\label{20b}
n_q^B \!&=&\! \tilde{\phi}_q^2 \tilde{n}_q^B \\
&&+ \tilde{\eta}_q^2 (1+
\tilde{n}_{-q}^B) + \sum_k \tilde{\psi}_{k,q}^2 \tilde{n}_{k+q}^c(1- 
\tilde{n}_k^c) ,\nonumber\\\label{20c}
d_k^c \!&=&\! \tilde{\alpha}_k \tilde{\alpha}_{k+Q} \tilde d_k^c \\
&&+ \sum_q \Big[ \tilde{\beta}_{k,q} \tilde{\beta}_{k+Q,q} 
(1+ \tilde{n}_q^B) + \tilde{\gamma}_{k,q} \tilde{\gamma}_{k+Q,q} \tilde{n}_q^B 
\Big] \tilde{d}_{k-q}^c\,. \nonumber
\end{eqnarray}
 Here, $\tilde{\alpha}_k, \tilde{\beta}_{k,q}, \cdots$ 
 denote the fully renormalized parameter values  at $\lambda=0$. Similarly, 
 $\tilde{n}_k^c, \tilde{n}_q^B$, and $\tilde d_k^c$ are 
expectation values defined with 
 $\tilde {H}$, i.e., 
 \begin{eqnarray}
\label{21a}
 \tilde{n}_k^c &=& \langle c_k^\dagger c_k^{} \rangle_{\tilde{H}}\,,\\ \label{21b}
  \tilde{n}_q^B &=& \langle B_q^\dagger B_q^{} \rangle_{\tilde{H}}\,,\\ \label{21c}
 \tilde d_k^c &=& \langle c_k^\dag c_{k+Q}^{} \rangle_{\tilde{H}}\,,  
  \end{eqnarray}
 where the fully renormalized Hamiltonian $\tilde{H}$
 is given by Eq.\eqref{17}.

\subsubsection{Dynamical correlation functions}

Let us consider the boson spectral function,
\begin{equation}
  \label{G3}
  C_q(\omega) =
  \frac{1}{2\pi\omega} \int_{-\infty}^{\infty}
  \left\langle [b_{q} (t), \; b_{q}^\dagger] \right\rangle \;
  e^{i\omega t} \, dt\,,
\end{equation}
and the two electronic one-particle spectral functions
\begin{eqnarray}
  \label{G4}
  A_{k}^{+}(\omega) &=& 
  \frac{1}{2\pi} \int_{-\infty}^{\infty}
  \left\langle c_{k} (t) \; c_{k}^\dagger \right\rangle \;
  e^{i\omega t} \,dt\,,\\
  A_{k}^{-}(\omega) &=& 
  \frac{1}{2\pi} \int_{-\infty}^{\infty}
  \left\langle c_{k}^\dagger \; c_{k}(t) \right\rangle \; 
  e^{i\omega t} \, dt \,. \label{G4b}
\end{eqnarray}
Here, 
$A_{k}^{+}(\omega)$ describes the creation of an electron with
momentum $k$ at time zero and its annihilation at time $t$ whereas in
$A_{k}^{-}(\omega)$ first an electron is annihilated. As it is well-known, 
$A_{k}^{+}(\omega)$ and $A_{k}^{-}(\omega)$ can be measured by inverse
photoemission and by photoemission. 

To evaluate Eqs. (\ref{G3})--(\ref{G4b}) within the PRM approach, we use
again that expectation values are invariant with respect to a unitary 
transformation under the trace. Thus, $C_q(\omega)$, $A_{k}^{+}(\omega)$, and
$A_{k}^{-}(\omega)$ can easily be computed if the bosonic and electronic
one-particle operators are transformed in the same way as the Hamiltonian. 
In this way we obtain 
\begin{widetext}
\begin{eqnarray}
  \label{G5}
  C_q(\omega) &=& 
  \frac{\tilde{\phi}_{q}^{2}}{\tilde{\omega}_{q} }
  \delta(\omega - \tilde{\omega}_{q}) + 
  \frac{\tilde{\eta}_{q}^{2}}{\tilde{\omega}_{-q}}
  \delta(\omega + \tilde{\omega}_{-q}) 
%\nonumber\\ &&+ 
+  \sum_{k} \tilde{\psi}_{k,q}^{2}
  \frac{\tilde{n}_k^c - \tilde{n}_{k+q}^c}{
    \tilde{\varepsilon}_{k+q} - \tilde{\varepsilon}_{k}
  }
  \delta(\tilde{\varepsilon}_{k+q} - \tilde{\varepsilon}_{k} - \omega), \\\label{A6-}
A_k^-(\omega) &=& \tilde{\alpha}_k^2  \tilde{n}_k^c \delta(\omega -
  \tilde{\varepsilon}_{k}) + \sum_q \Big[ \tilde{\beta}_{k,q}^2
  \left(1 + \tilde{n}_q^B\right) \tilde{n}_{k+q}^c \delta(\omega +
  \tilde{\omega}_q - \tilde{\varepsilon}_{k+q}) 
+ \, \tilde{\gamma}_{k,q}^2
  \tilde{n}_q^B \tilde{n}_{k-q}^c \delta(\omega -
  \tilde{\omega}_q - \tilde{\varepsilon}_{k-q}) \Big],
\\\label{A6+}A_k^+(\omega) &=& \tilde{\alpha}_k^2  (1 - \tilde{n}_k^c) \delta(\omega -
  \tilde{\varepsilon}_{k}) \nonumber \\
&&+ \sum_q \Big[ \tilde{\beta}_{k,q}^2
  \tilde{n}_q^B (1 - \tilde{n}_{k+q}^c) \delta(\omega +
  \tilde{\omega}_q - \tilde{\varepsilon}_{k+q}) 
+ \, \tilde{\gamma}_{k,q}^2
  (1 + \tilde{n}_q^B) (1 - \tilde{n}_{k-q}^c) \delta(\omega -
  \tilde{\omega}_q - \tilde{\varepsilon}_{k-q}) \Big], 
\end{eqnarray}
\end{widetext}
where terms with two bosonic creation or annihilation operators have been
neglected. The $\tilde{\phi}_{q}$, $\tilde{\eta}_{q}$, and 
$\tilde{\psi}_{k,q}$ are the zero-$\lambda$ coefficients taken from the 
evaluation of Eqs.\eqref{phi}-\eqref{psi}. 

Let us emphasize that the expressions \eqref{G3}, \eqref{G4} and \eqref{G4b} 
fulfill the sum rules, 
\begin{equation}
  \label{G5b}
\int_{-\infty}^\infty \, d\omega \, \omega \, C_q(\omega) = 1
\end{equation}
and
\begin{equation}
\int_{-\infty}^{\infty} d\omega \, [A_{k}^{+}(\omega)+A_{k}^{-}(\omega)] =
1\,,
\end{equation}
respectively, which also hold if Eq.~(\ref{G5}) for $C_{q}(\omega)$ and 
the corresponding expressions for 
$A_{k}^{+}(\omega)$ and $A_{k}^{-}(\omega)$ are inserted. 

\newpage

\subsubsection{Numerical analysis}
%\section{Numerical results and discussion} 
The set of renormalization equations \eqref{15}--\eqref{16b}  
and \eqref{alpha}--\eqref{psi} has to be solved numerically. 
To this end, we choose some initial values for the expectation
values entering the renormalization equations. Using this set of quantities, 
the numerical evaluation starts from the cutoff $\bar{\lambda}$ of the
original model $H$ and proceeds step by step to $\lambda = 0$. For
this procedure we consider a lattice of $N = 500$ sites in one 
dimension. The width of the energy shell $\Delta \lambda$ was taken to
be somewhat smaller than the typical smallest energy spacing of the
eigenstates of $H_{0,\lambda}$. For $\lambda = 0$, the Hamiltonian and
the one-particle operators are fully renormalized. The case $\lambda =
0$ allows the re-calculation of all expectation values, and the
renormalization procedure starts again with the improved expectation
values by reducing again the cutoff from $\bar{\lambda}$ to $\lambda =
0$. After a sufficient number of such cycles, the expectation values
are converged and the renormalization equations are solved
self-consistently. Convergence is assumed to be achieved if all
quantities are determined with a relative error less than
$10^{-5}$. The dynamical correlation functions~\eqref{G5}--\eqref{A6+}
are evaluated using a broadening in energy space that is equal 
to $\Delta \lambda$.
\section{Ground-state properties}
\subsection{DMRG phase diagram}
In order to classify the PRM  ground-state and spectral properties 
given below, we first present in Fig.~\ref{fig1} a refined version of the 
DMRG ground-state phase diagram of the half-filled fermion-boson 
model~\eqref{hem}. Here the phase boundary, separating the insulating phase 
with CDW long-range order from the metallic TLL phase
in the $\Lambda$--$\Omega$ plane, was obtained from the
$N \to \infty$ extrapolated values of the Luttinger liquid parameter
$K_\rho$ and the single particle (charge) gap $\Delta_c$~\cite{EHF09,SEprivate}.
In the limit of large $\Omega$, the background fluctuations, associated
with any particle hop, are energetically costly. As a result the motion of 
the particle is hindered and charge ordering becomes favorable if 
$\Lambda$, describing the ability of the background to relax, is sufficiently 
low. By contrast, for large $\Lambda$ ($\Lambda >\Lambda_c(\Omega =\infty) 
\simeq 0.1588$), we find metallic behavior for all $\Omega$. 
In the limit of small $\Omega$, the rate of bosonic fluctuations 
($\propto \Omega^{-1}$) is high. Then, in no way, correlations 
emerge within the background medium. The DMRG results suggest that 
for $\Lambda =0$, i.e. when the relaxation channel is closed, the 
ground state is nevertheless metallic below a finite critical boson energy 
$\Omega_c(\Lambda =0)$.  Let us re-emphasize that coherent 
particle hopping is possible 
even when $\Lambda =0$, due to a six-step 
vacuum-restoring hopping process~\cite{AEF07},
\begin{equation}\label{6step}  
R_{i,i+2}^{(6)}=L^\dagger_{i+2} L^\dagger_{i+1} R^\dagger_i L_{i+2} R_{i+1} R_i
\end{equation} 
with
$R_i^{\dagger}=c_{i}^{\dagger}c_{i+1}^{}b_i^{}$ and 
$L_i^{\dagger}=c_{i}^{\dagger}c_{i-1}^{}b_i^{}$. $R_{i,i+2}^{(6)}$ leads to 
an `effective' (coherent) next nearest-neighbor transfer. 
%`$c_{i+2}^{\dagger} c_i^{}$'. 
\begin{figure}[hbt]
\includegraphics[width=0.9\linewidth]{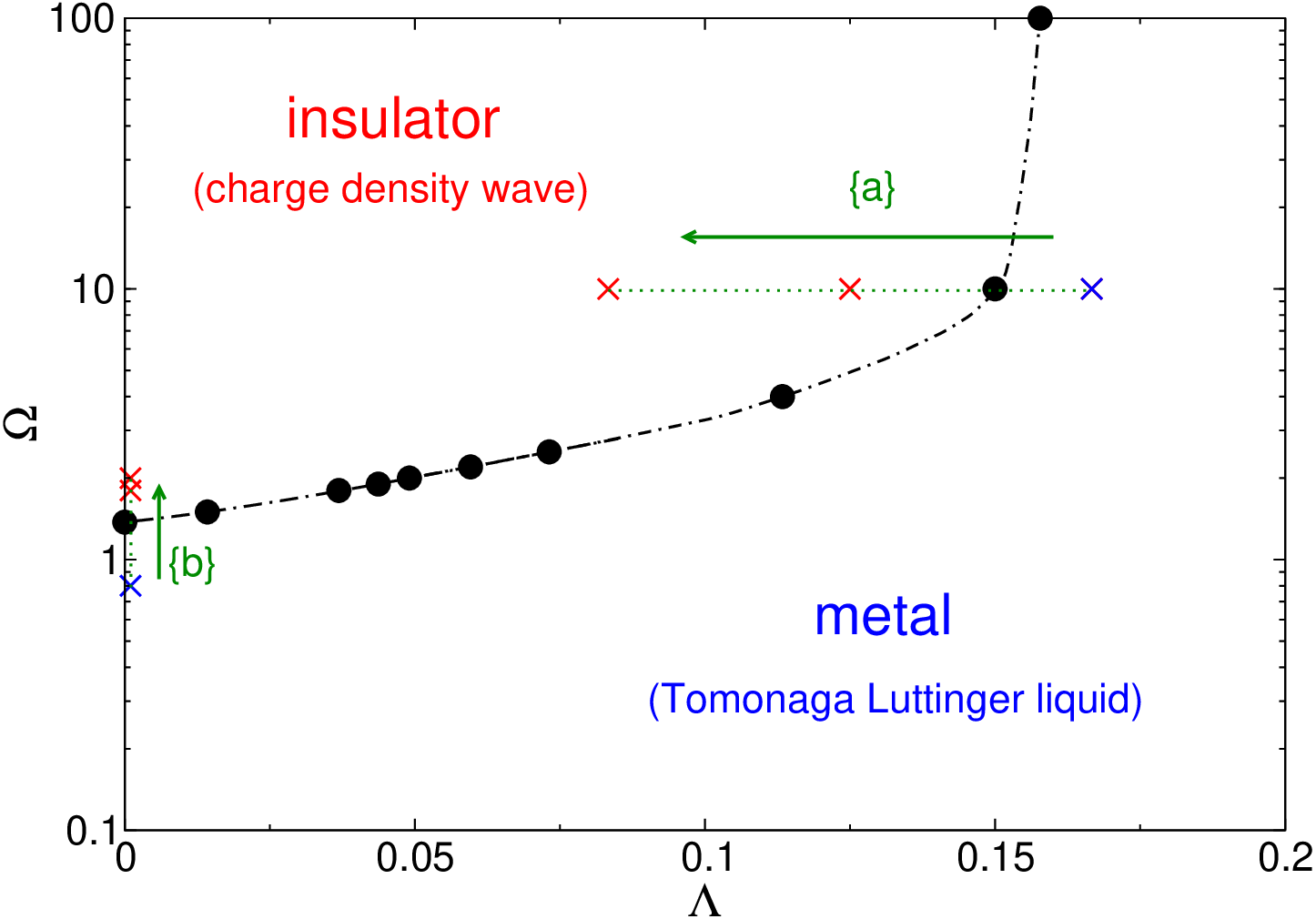}
%pd-Edwards-1.1.eps
\caption{(Color online)  DMRG phase diagram of the two-channel fermion-boson
transport model~\eqref{hem} [see also Eq.~\eqref{1}] for the 1D 
half-filled band case (the dot-dashed line is a 
guide to the eye).  The crosses mark the parameter values used within our 
PRM calculation when passing the TLL$\to$CDW transition at fixed
$\Omega=10$ \{a\} and $\Lambda = 0.001$ \{b\} (green arrows), respectively.  
\label{fig1}}
\end{figure}

Concerning the nature of the metal-insulator quantum phase transition,  
it is a moot point, whether the TLL-CDW crossover in the half-filled 
fermion-boson model~\eqref{hem}, taking place at relatively large $\Omega$, 
bears some resemblance to the usual Peierls transition in the spinless 
fermion Holstein model~\cite{BMH98,SHBWF05,HWBAF06,EF09b}. In this regard 
the question of boson softening will certainly be of importance.    

\subsection{CDW order parameter}
To analyze the nature of the metal-insulator transition of~\eqref{hem} 
in more detail, we calculate in the following a set of characteristic 
quantities by the PRM for the (1D half-filled) infinite system. Thereby 
we cross the TLL$\to$CDW transition in the following figures 
at fixed $\Omega$ [panels (a)]
and $\Lambda$ [panels (b)] (cf. Fig.~\ref{fig1} lines \{a\} and \{b\}, 
respectively). As for the half-filled
Holstein model~\cite{SHB09},
%,SSykoraprivate}, 
the CDW structure of the
insulating state shows up in the correlation function 
$d_k^c= \langle c_k^\dag c_{k+Q}^{}\rangle$, which can be considered as 
CDW order parameter.
\begin{figure}[hbt]
\includegraphics[width=0.9\linewidth]{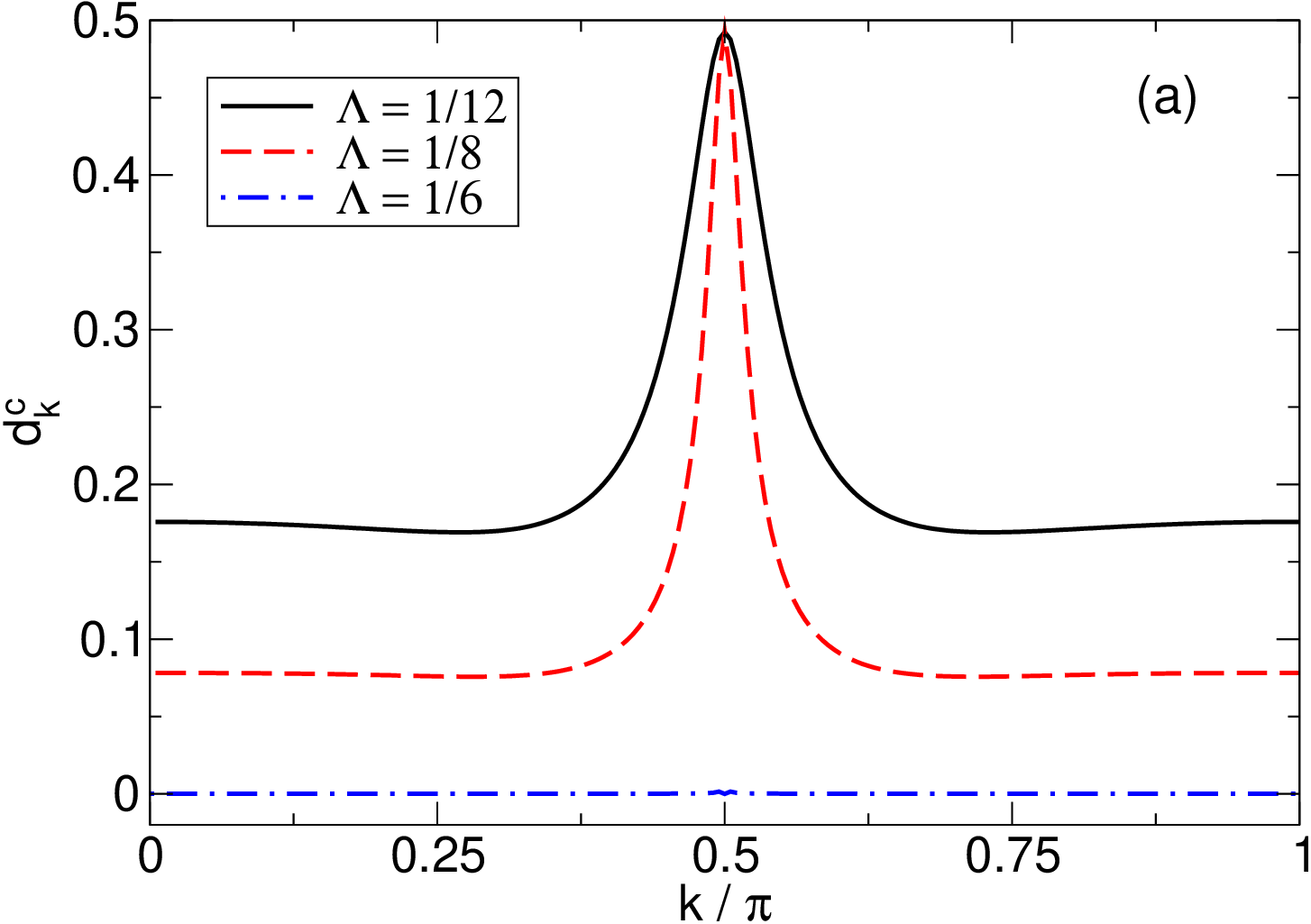}\\[0.1cm]
%erw_d_c_w01.eps
\includegraphics[width=0.9\linewidth]{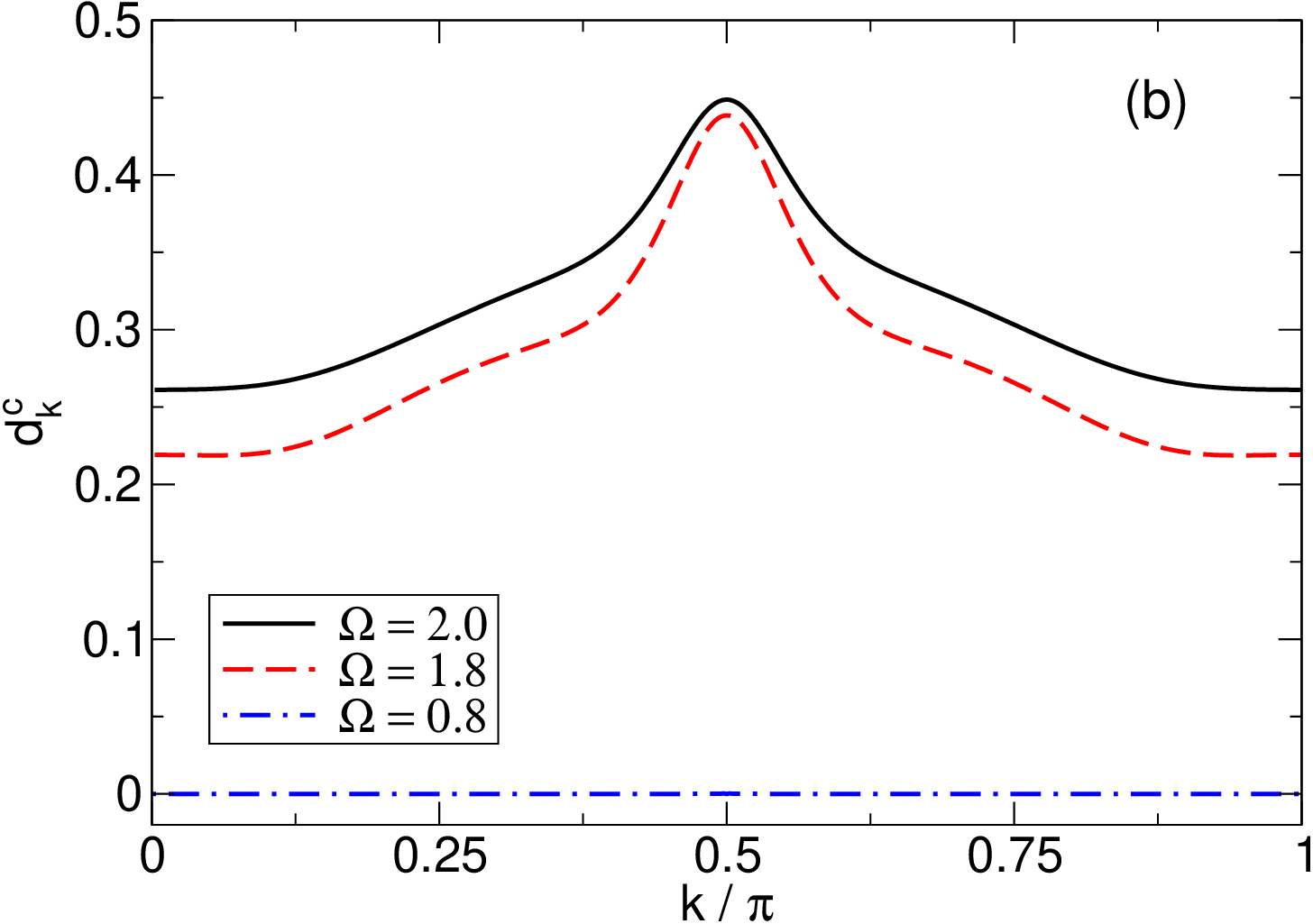}
%erw_d_c_l1000.eps
\caption{(Color online) Zero-temperature expectation value $d_{k}^c$, 
indicating CDW order in the half-filled fermion-boson model~\eqref{hem}, 
at $\Omega  = 10$ [upper panel (a)] and $\Lambda =0.001$ [lower panel (b)].}
\label{fig2}
\end{figure}

Figure~\ref{fig2} displays the variation of this 
expectation value when the wave vector $k$ runs through the half  
1D Brillouin zone. Obviously we have $d_k^c=0$ in the metallic phase
(blue dot-dashed line). Entering the CDW state $d_k^c$  acquires 
finite values, whereby the maximum of $d_k^c$ is at $k=\pi/2$ 
(for this case $Q=\pi$ connects
both Fermi momenta $k_F =\pm \pi/2$). If the charge order is  
perfect (the particles are localized in an A-B structure without 
any charge fluctuations, i.e. the lower and upper bands are flat), we 
find $d_{k}^c=1/2$ for all $k$. This tendency becomes apparent by 
comparing the results obtained in the CDW phase for different 
$\Lambda$ (cf. upper panel $\Lambda =1/8,\;1/12$ and lower
panel $\Lambda =1/1000$; recall that $\Lambda$ describes the ability
of the background to relax.) 
\subsection{Fermion dispersion and quasiparticle weight}
\begin{figure}[b]
\includegraphics[width=0.9\linewidth]{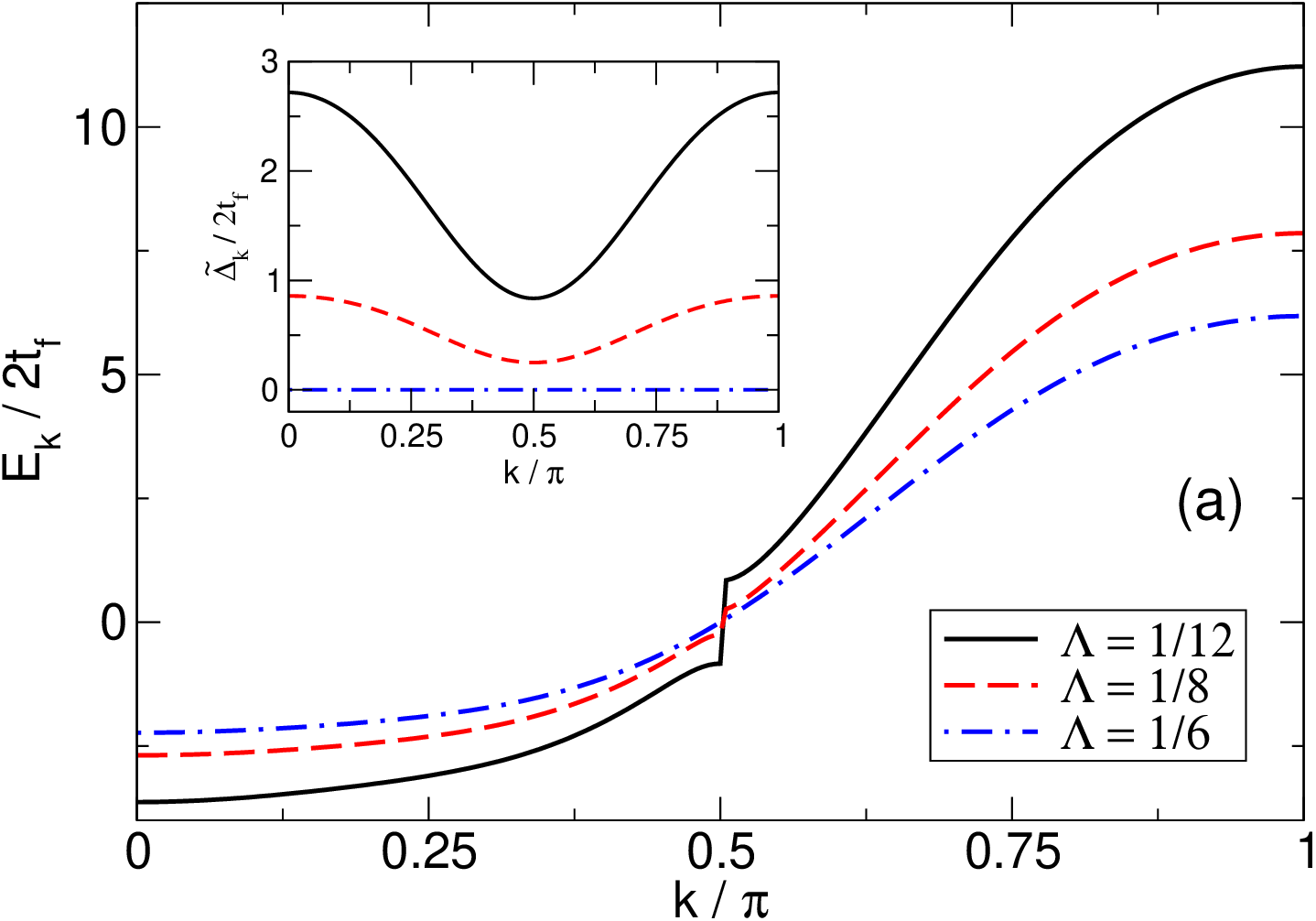}\\[0.1cm]
%ren_e_c_w01.eps
\includegraphics[width=0.9\linewidth]{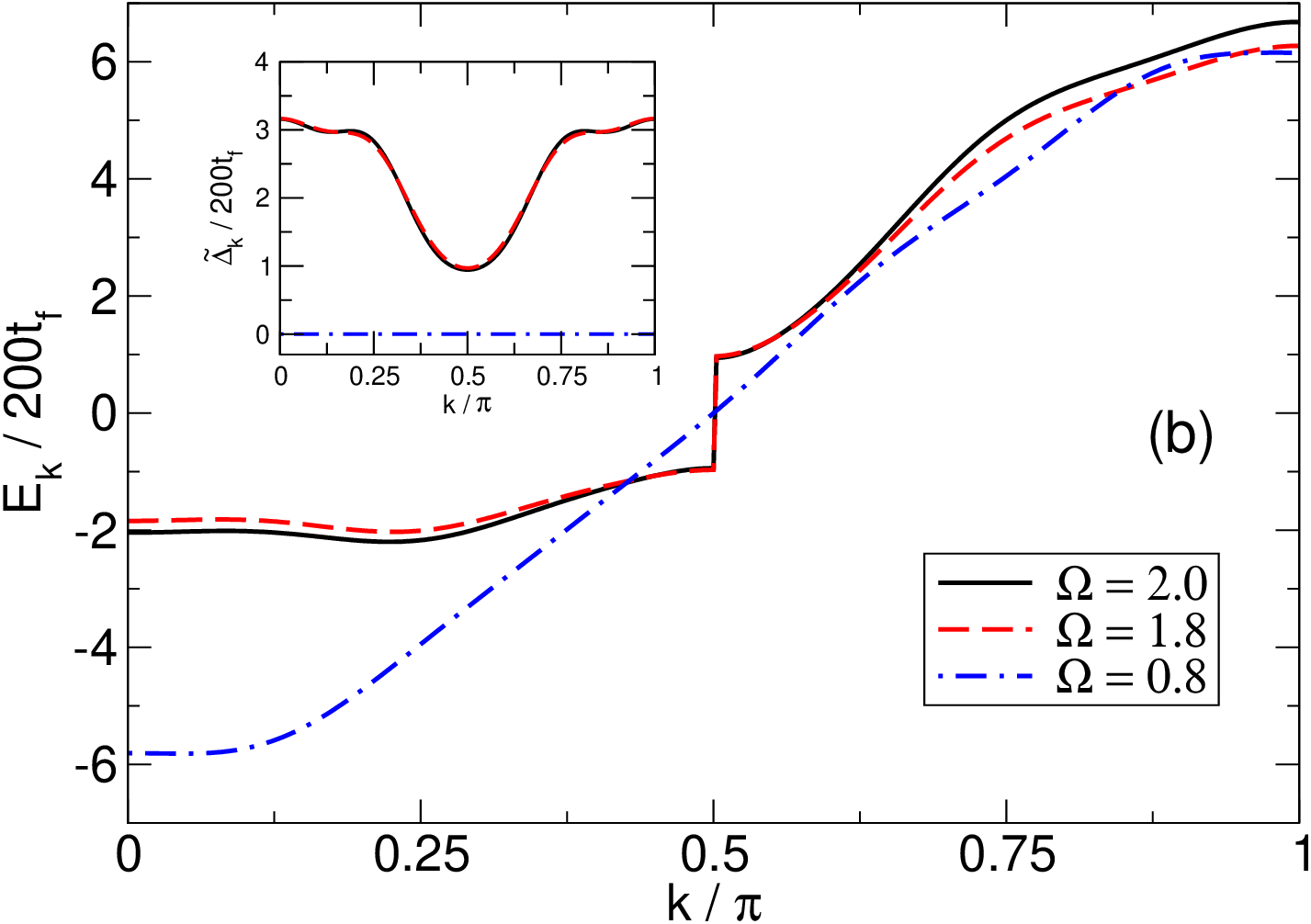}
%ren_e_c_l1000.eps
\caption{(Color online) Dispersion of the occupied lower (empty upper) 
quasiparticle band $E_k$ 
%= {(+)\atop -}
%[\tilde{\varepsilon}_k^2 + 
%\tilde{\Delta}_k^2]^{1/2}$ 
with $k\leq \pi/2$ ($k > \pi/2$) for 
$\Omega =10$ [panel (a)] and $\Lambda =0.001$ 
[panel (b)] (the Fermi energy sets the zero-point of energy). 
Note that band energies are differently scaled by $2t_f$ and $200 t_f$
in (a) and (b). The insets show the $k$-dependence 
of $\tilde{\Delta}_k$.}
\label{fig3}
\end{figure}

Next we investigate the renormalization of the fermionic  
band structure, 
\begin{equation}
E_k=
\frac{\tilde{\varepsilon}_k+\tilde{\varepsilon}_{k+Q}}{2}
\pm \sqrt{\Big(\frac{\tilde{\varepsilon}_k-\tilde{\varepsilon}_{k+Q}}{2}\Big)^2
+|\tilde{\Delta}_k|^2} \,,
\end{equation}
see Fig.~\ref{fig3}. 
In the {\it metallic regime}, of course, 
there is no gap at the Fermi energy (Fermi vector $k_F=\pi/2$), and 
$\tilde{\Delta}_k$, given in the inset, is zero for all $k$.
While for $2\Lambda \gg \Omega$ the free transport channel ($\propto t_f$) 
dominates even when $\Omega$ is large (see Eq.\eqref{t_f_L}), 
the bosonic degrees of freedom
will strongly affect the transport for small $\Lambda$. As a consequence,
`coherent' transport takes place on a strongly reduced energy scale only 
(we have $t_f/t_b = 1/30$ [$t_f/t_b =1/400$]  
for the blue dot-dashed line in panel~(a)~[(b)]).

\begin{figure}[t]
\includegraphics[width=0.9\linewidth]{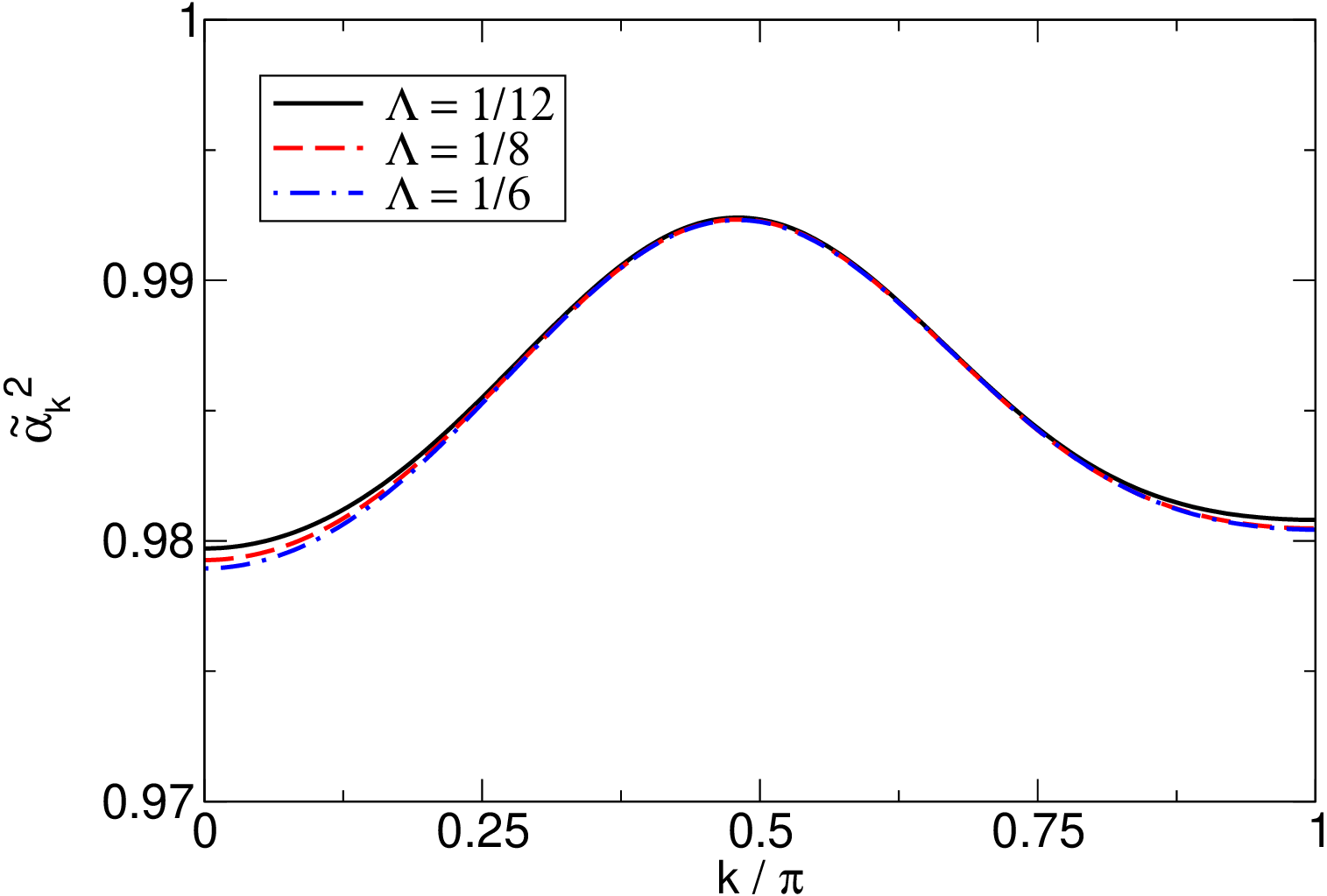}\\[0.1cm]
%ren_e_alpha_w01.eps
\hspace*{0.4cm}\includegraphics[width=0.84\linewidth]{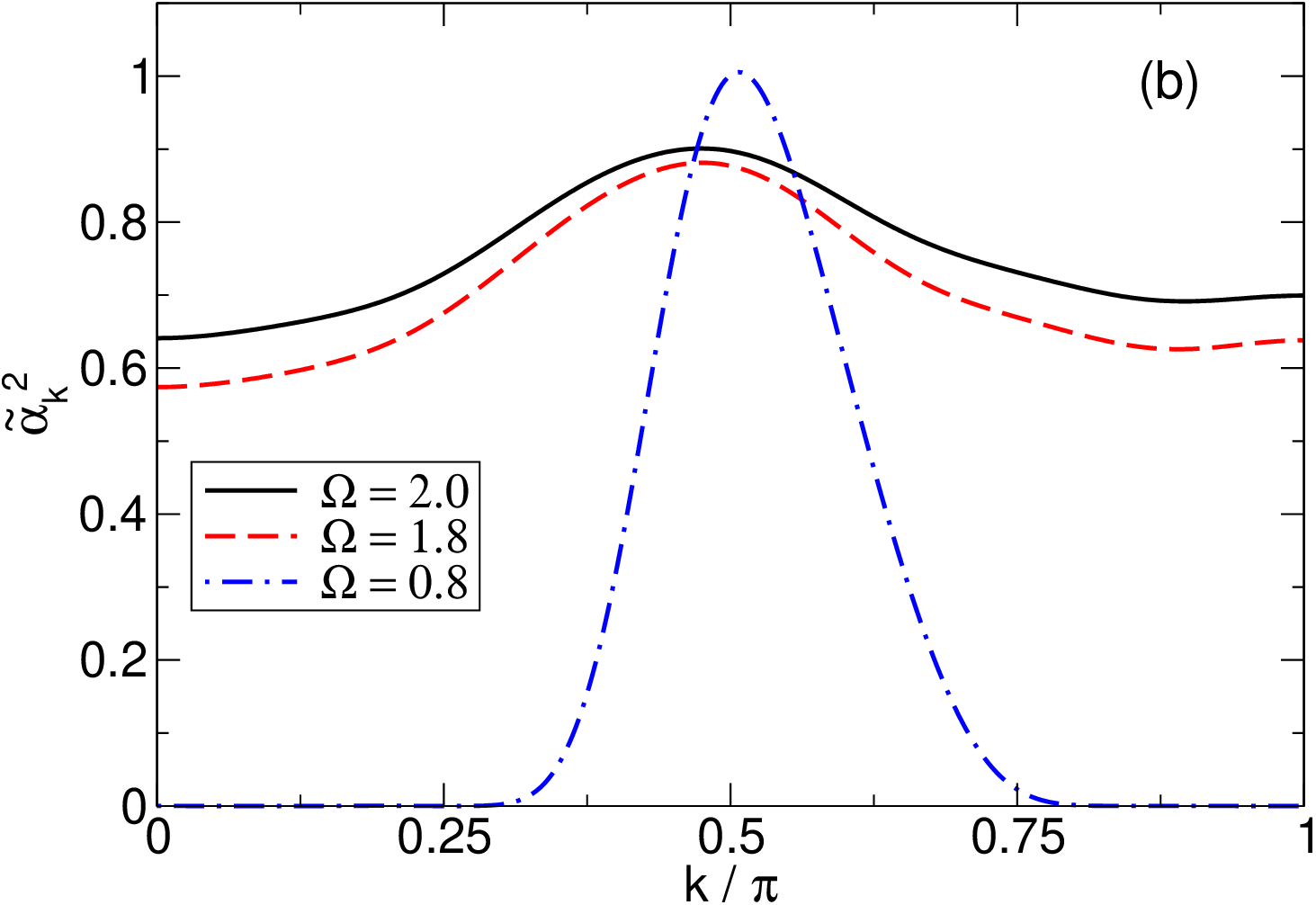}
%ren_e_alpha_l1000.eps
\caption{(Color online) Spectral weight, $\tilde{\alpha}_k^2$, of the 
coherent (quasiparticle) part of the electronic one-particle 
spectral function~\eqref{A6-}. We have $\Omega =10$ in panel (a) and 
$\Lambda=0.001$ in panel (b). Notations are as in Fig.~\ref{fig3}.}
\label{fig4}
\end{figure}

The coefficient $\tilde{\alpha}_k^2$, depicted in 
Fig.~\ref{fig4}, gives the weight of the corresponding coherent part of 
the single-particle spectral function~\eqref{A6-}.  
$A^{-}_k(\omega)$ can by probed by angle-resolved photoemission experiments. 
At very large $\Omega$ (and small $\Lambda$), the particles will 
solely move by the above mentioned six-step process~\eqref{6step}. 
Then the  resulting 
`quasiparticle weight', $\tilde{\alpha}_{k}^2$, is nearly one 
[see Fig.~\ref{fig4}~(a)], and shows a very weak $k$-dependence. 
For small $\Omega<\Omega_c$  we enter the fluctuation-dominated regime
and the nature of the metallic state changes noticeably. In accordance
with recent dynamical DMRG data for the single-particle spectra,
which show that the absorption spectrum is over-damped near $k=0,\,\pi$  
because of intersecting bosonic excitations,  we find 
$\tilde{\alpha}_k^2\sim 1$ in the vicinity of $k_F$ only 
[cf. Fig.~3 in Ref.~\onlinecite{EF09b} and Fig.~\ref{fig4}~(b), 
blue dot-dashed line].        

In the {\it insulating regime}, the renormalized band structure
$E_k$ is gaped (see Fig.~\ref{fig3}, dashed and solid lines). 
The inset in
Fig.~\ref{fig3} clearly shows an increase of the gap at $k = k_F$ as $\Lambda$ 
gets smaller. Note that
the size of the gap is equal to $2 \tilde{\Delta}_{k_F}$. 
While $\tilde{\Delta}_k$ is symmetric around $k=\pi$, $E_k$ 
is not. The reason is that doping a perfect CDW, states with one 
particle removed are connected by the six-step hopping process~\eqref{6step},  
whereas 
a two-step hopping process relates states with an additional 
particle~\cite{WFAE08}. In this way the collective particle-boson dynamics
leads to a more pronounced flattening of the coherent band for $k<k_F$,
i.e., the widths of the highest photoemission and lowest inverse 
photoemission band differ~\cite{WFAE08,EF09b}. It is encouraging that 
our analytic PRM approach reproduces this non-trivial correlation-induced 
(mass-) asymmetry. Let us emphasize that the $\tilde{\alpha}_k^2$ 
given in Fig.~\ref{fig4} for the CDW case (dashed and solid curves) 
belong to the highest photoemission band in the whole interval $[0,\pi]$ 
(the corresponding $E_k$ is not depicted in the region $\pi/2 < k \leq \pi$ 
in Fig.~\ref{fig3}). Compared to the metallic phase the spectral
weight of the lower CDW band is significantly changed for 
intermediate-to-small boson frequencies only.       

\subsection{Boson dispersion and occupation numbers}
\begin{figure}[t]
\includegraphics[width=0.9\linewidth]{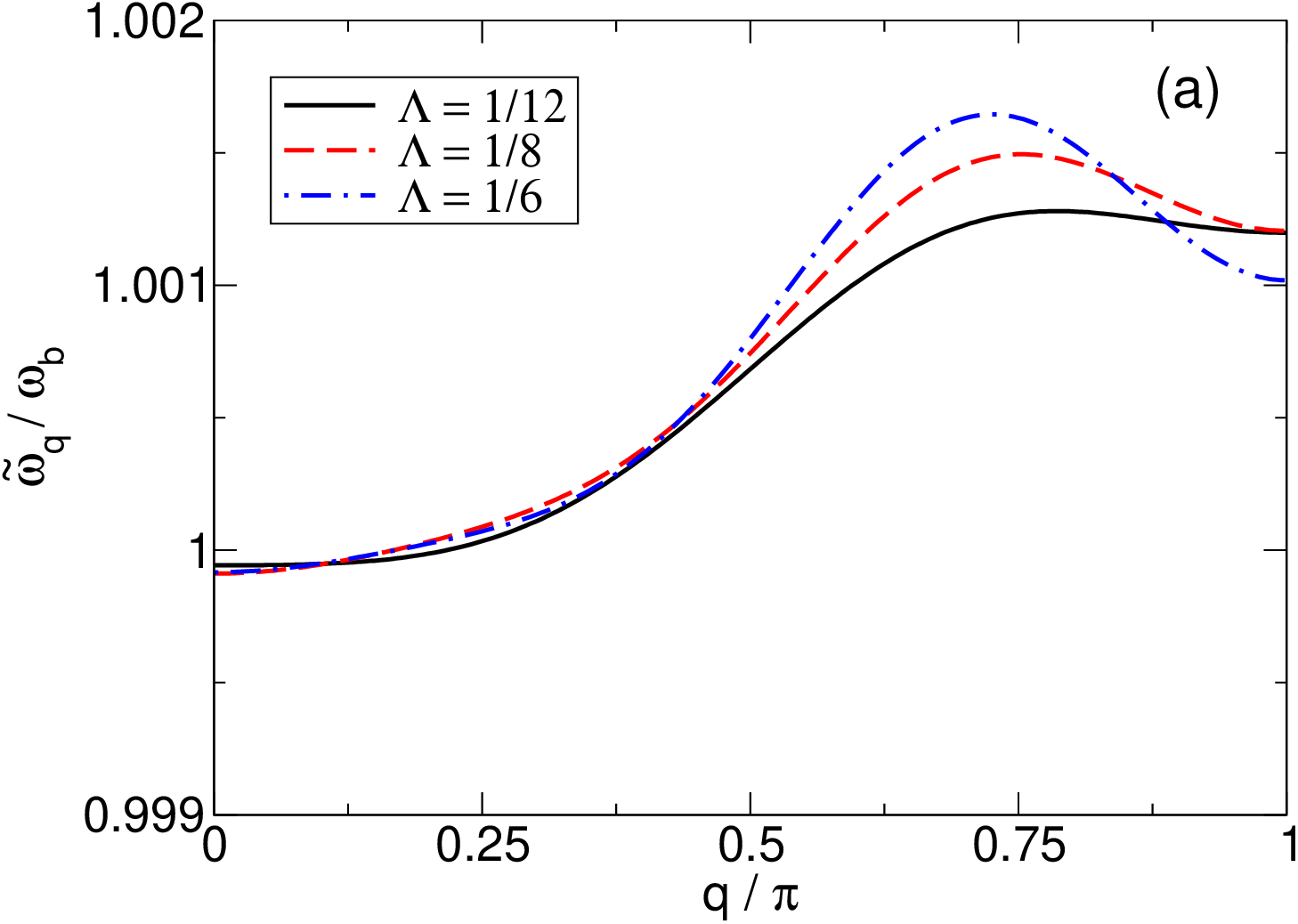}\\[0.1cm]
%ren_e_w_w01.eps
\hspace*{0.2cm}\includegraphics[width=0.86\linewidth]{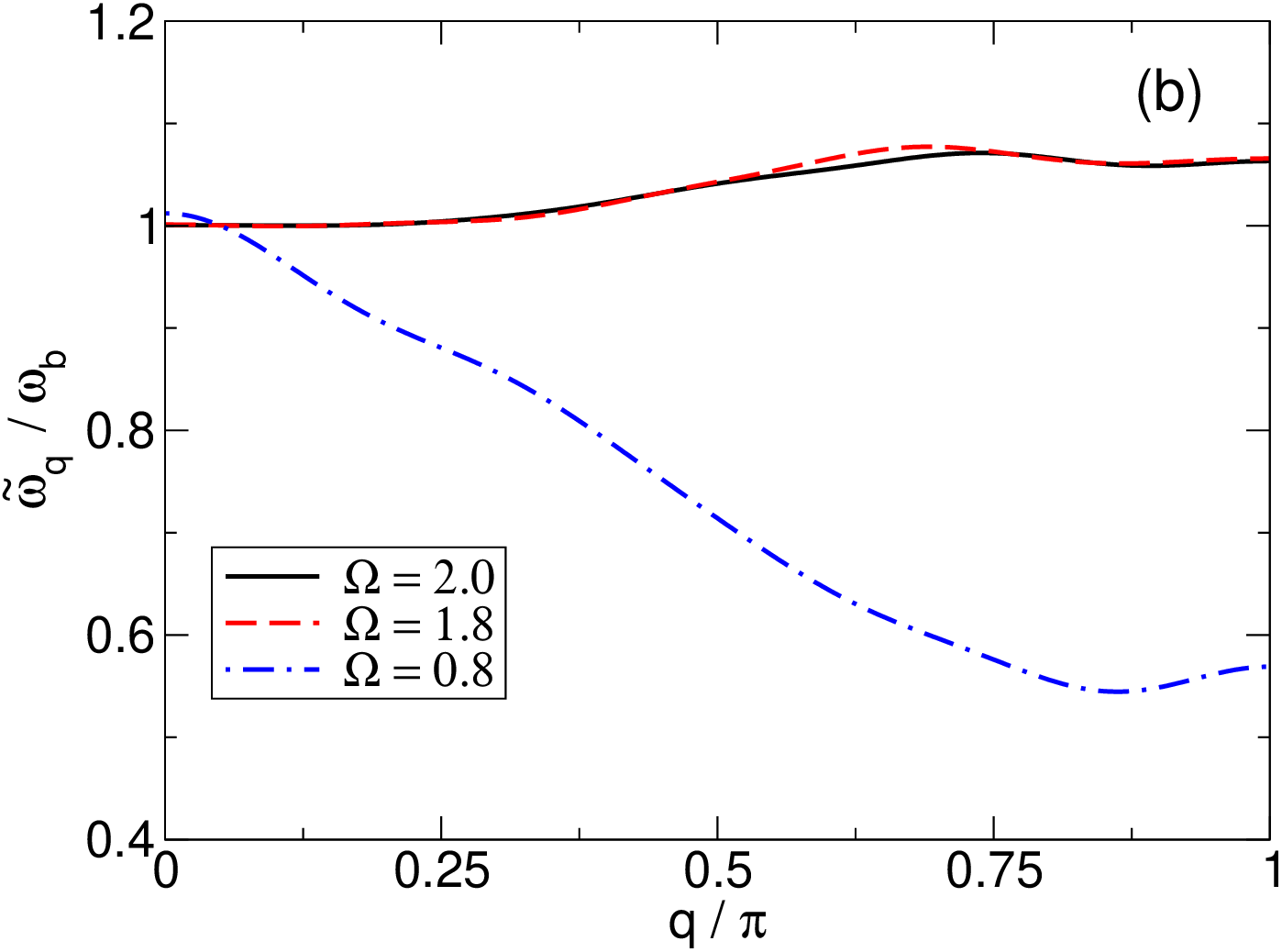}
%ren_e_w_l1000.eps
\caption{(Color online) Renormalized boson frequencies 
$\tilde{\omega}_q$ for the fermion-boson transport model~\eqref{hem}
with $\Omega = 10$ [panel (a)]
and $\Lambda =0.001$ [panel (b)].}
\label{fig5}
\end{figure}

The Einstein bosons, describing excitations of the background, 
gain a dispersion owing to the coupling to the fermions. The 
renormalization of the boson dispersion, $\tilde{\omega}_q/\omega_b$, 
is displayed in Fig.~\ref{fig5}. It is rather weak for large $\Omega$ 
in both the metallic and insulating states. For smaller boson frequencies,  
we find a strong renormalization in the TLL phase (up to 50\% for 
$\Omega =0.8$) at larger momenta [see dot-dashed curve in panel (b)]. 
This is in accordance with the over-damped single-particle excitations 
observed in the ARPES spectra~\cite{EF09b} and, of course, also shows
up in the depletion of $\tilde{\alpha}_k^2$ away from $k=\pi/2$ 
[see Fig.~\ref{fig4}~(b)].  

Most notably, for the CDW state, we observe a hardening of the boson modes 
near the Brillouin zone boundary. This holds in the whole $\Omega$ region 
and means that the TLL$\to$CDW transition is unlike the usual displacive 
Peierls transition which, in general, is accompanied by the softening of 
the $q=\pi$ boson (phonon)~\cite{SHBWF05,HWBAF06}.
In our case, the CDW state is driven 
by the stiffness 
of the background, being most pronounced at large $\Omega$ and small 
$\Lambda$. By contrast, when $\Omega$ is small, i.e. the background 
readily fluctuates, the kinetic energy part will naturally overcompensate
any potential energy gain by charge ordering. Another reason for the 
absence of boson softening might be the particular form of the fermion-boson 
interaction. As can be seen from the Fourier transformed Hamiltonian~\eqref{5},
the fermion-boson coupling vanishes for $k=\pm \pi/2$, i.e., precisely 
for the Fermi momenta of the half-filled band case. 
\begin{figure}[t]
\hspace*{0.05cm}\includegraphics[width=0.88\linewidth]{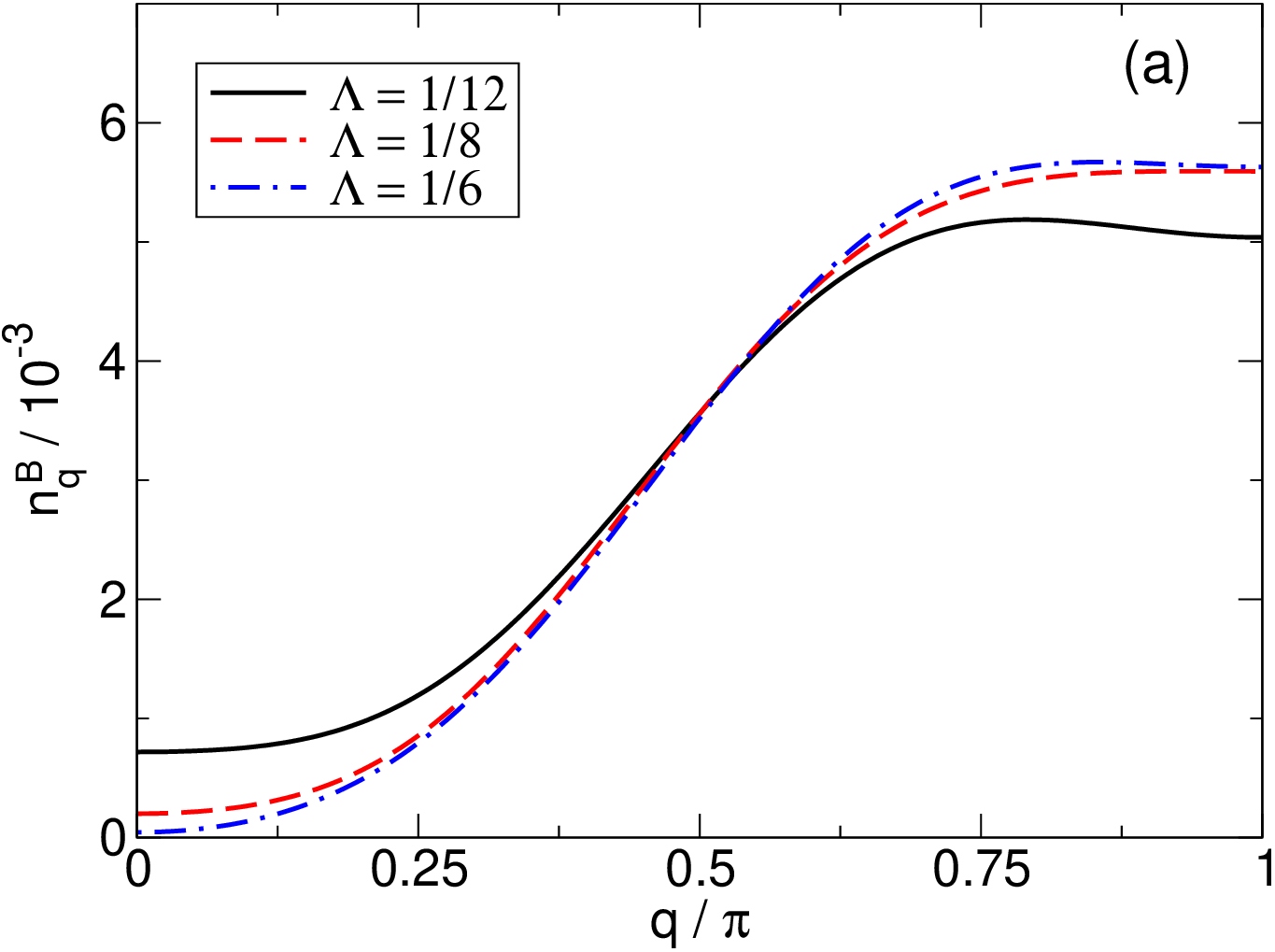}\\[0.1cm]
%erw_e_b_w01.eps
\includegraphics[width=0.9\linewidth]{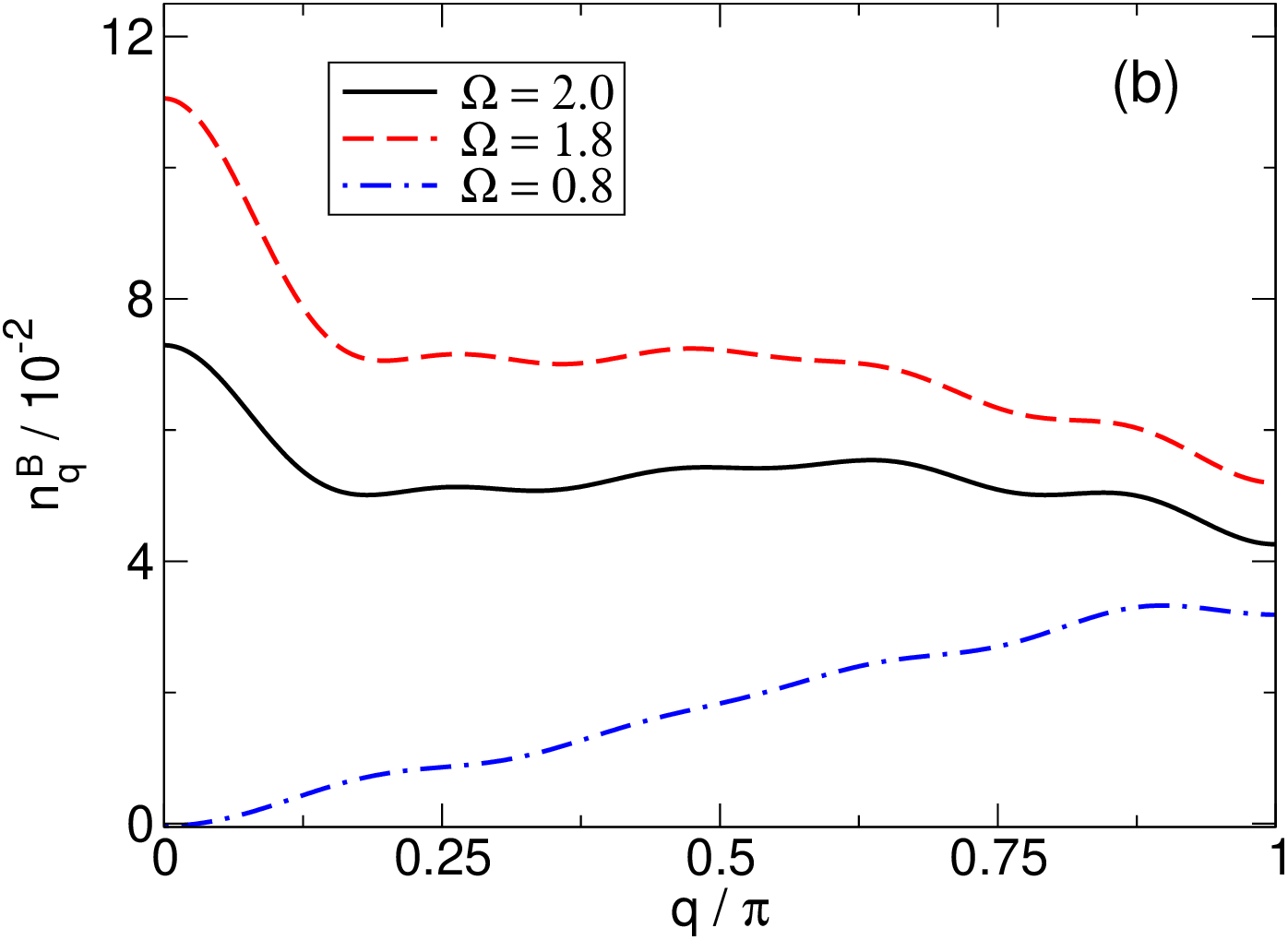}
%erw_e_b_l1000.eps
\caption{(Color online) Transformed boson expectation value 
$n_q^B=\langle B^\dagger_q B^{}_q \rangle$ at $\Omega= 10$ (a) 
and $\Lambda =0.001$ (b).}
\label{fig6}
\end{figure}

It may be worthwhile to demonstrate that our PRM approach has the
advantage that all features of the results for $\tilde{\omega}_q$ and all 
other renormalized quantities can easily be understood on the basis of the 
former renormalization equations. For simplicity we shall restrict ourselves
to the renormalization of $\tilde{\omega}_q$ in the case of large $\Omega$.
In this regime, from Eq.\eqref{omeg} one may point out the stiffening 
of the boson modes. Since the boson energy $\omega_b$ is much larger than 
the electronic bandwidth, for all $\lambda$ a positive energy denominator 
$(\varepsilon_{k,\lambda} - \varepsilon_{k+q,\lambda} +\omega_{q, \lambda})$
is obtained. Nevertheless, in the $k$ sum on the right hand side of 
Eq.\eqref{omeg} there are as many negative as positive terms due to the
factor $(n_k^c - n_{k+q}^c)$. Since from $(n_k^c - n_{k+q}^c)<0$ it follows
that $(\varepsilon_{k,\lambda} - \varepsilon_{k+q,\lambda})>0$, the negative 
terms have larger energy denominators and are always smaller than the positive 
terms. The resulting renormalization of $\tilde{\omega}_q$ is therefore 
positive for all $q$ values and largest for $q=\pi$ due to the smallest 
energy denominator. Furthermore, since $g_k \propto t_b$ and $\omega_b$ is
large the renormalization contributions in Eq.~\eqref{omeg} are of the order
of $t_b^2 / \omega_b = \omega_b / \Omega^2 \ll \omega_b$ which gives rise
to the weak dispersion of $\tilde{\omega}_q$ observed in Fig.~\ref{fig5}.
For smaller values of $\Omega$ the bosonic and fermionic energy values in
the denominator of Eq.~\eqref{omeg} can become comparable which immediately 
leads to a strong dispersion of $\tilde{\omega}_q$ (see dot-dashed
curve in Fig.~\ref{fig5} and solid curve in Fig.~\ref{fig8}).

Figure~\ref{fig6} gives the ($q$-resolved) boson occupation numbers. 
As one can see from Eq.~\eqref{20b}, this quantity for $T=0$ acquires 
finite values solely by coupling to fermionic degrees of freedom. 
Note that the first term in Eq.~\eqref{20b} vanishes for $T=0$.
We see that the formation of the CDW state is accompanied by a finite
occupation value of the $q=0$ boson mode, which is about two orders of 
magnitude larger if one compares $n_0^B$ for the CDWs established at 
$\Omega=1.8$ and  $\Omega=10$, respectively. Different from the 
Holstein-model CDW (Peierls) phase~\cite{SHBWF05}, the CDW phase of the 
half-filled fermion-boson transport model~\eqref{hem} is always a
few-boson state however. Referring to this, our PRM results corroborate 
previous small cluster exact diagonalization data~\cite{WFAE08}.
As can be seen from the third term of Eq.~\eqref{20b}, bosons having 
finite momentum give rise to an effective fermion interaction on neighboring 
sites.   

%\subsection{Boson spectral function in the metallic phase}
\section{Spectral properties}
While the fermionic single-particle spectral function 
of the transport model~\eqref{hem} was previously calculated 
for finite clusters by exact diagonalization~\cite{WFAE08} and dynamical 
DMRG~\cite{EF09b} techniques, the spectral response of the bosons   
has not been studied so far. The PRM allows to investigate 
the interrelation between fermion and boson dynamics by computing  
the boson spectral function, $C_q(\omega)$, according to Eq.~\eqref{G5},   
for the 1D infinite system~\cite{SHB06}. 
\begin{figure*}[h]
\includegraphics[width=0.42\linewidth]{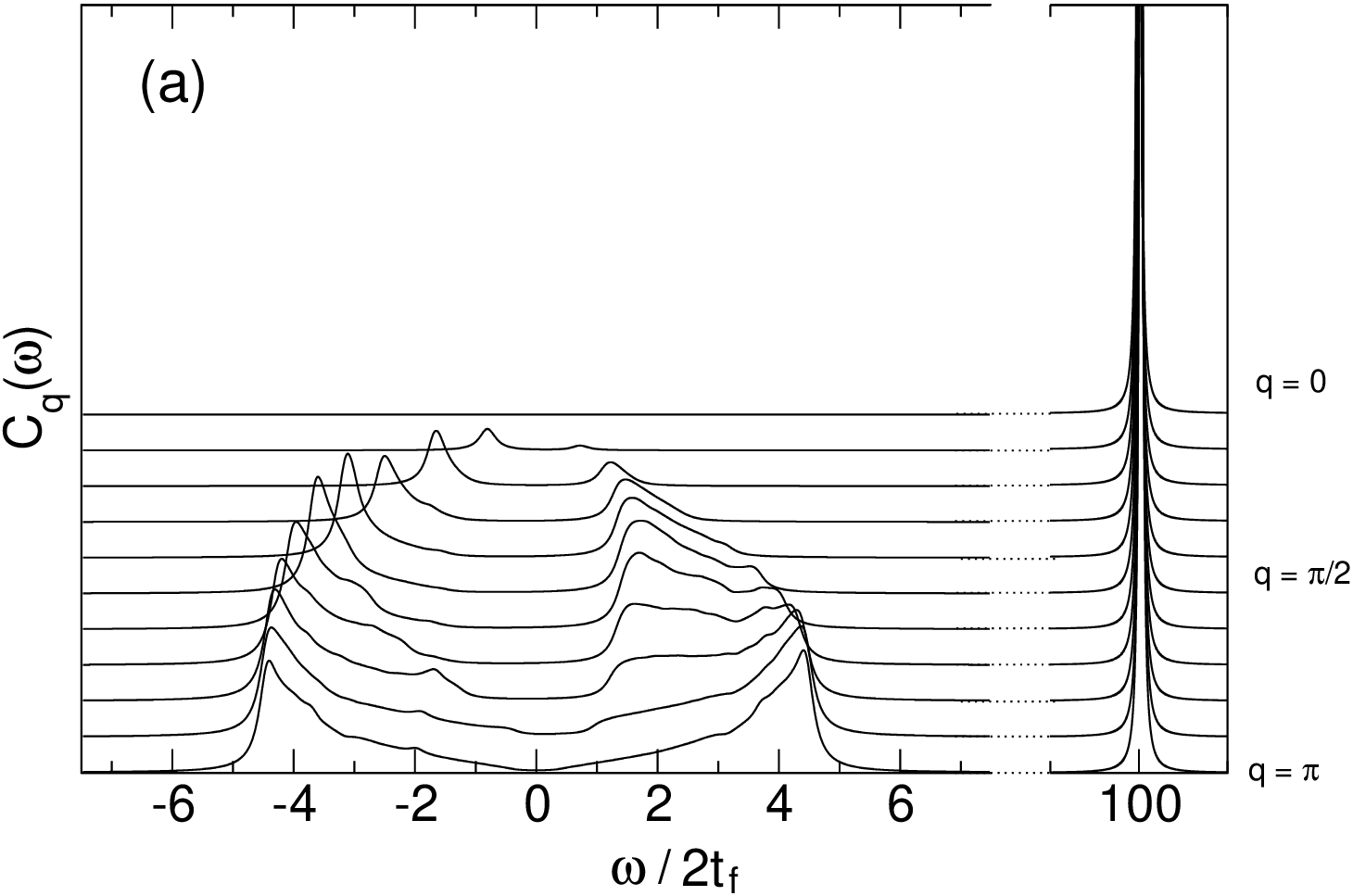}\hspace*{0.2cm}
\includegraphics[width=0.41\linewidth]{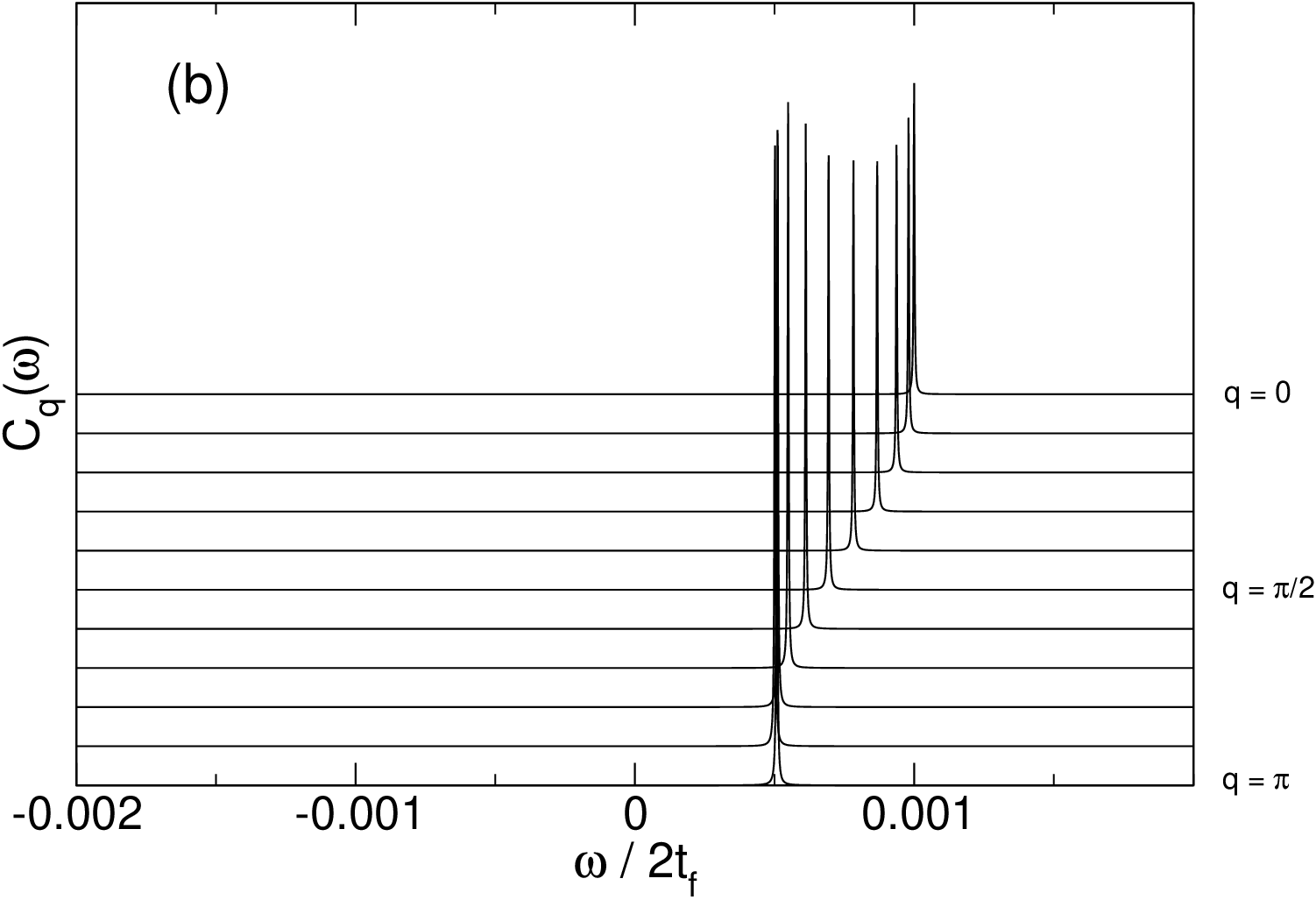}\\[0.2cm]
%B_q_w01_l4_n025.eps%B_q_w001_tb03.eps
\includegraphics[width=0.42\linewidth]{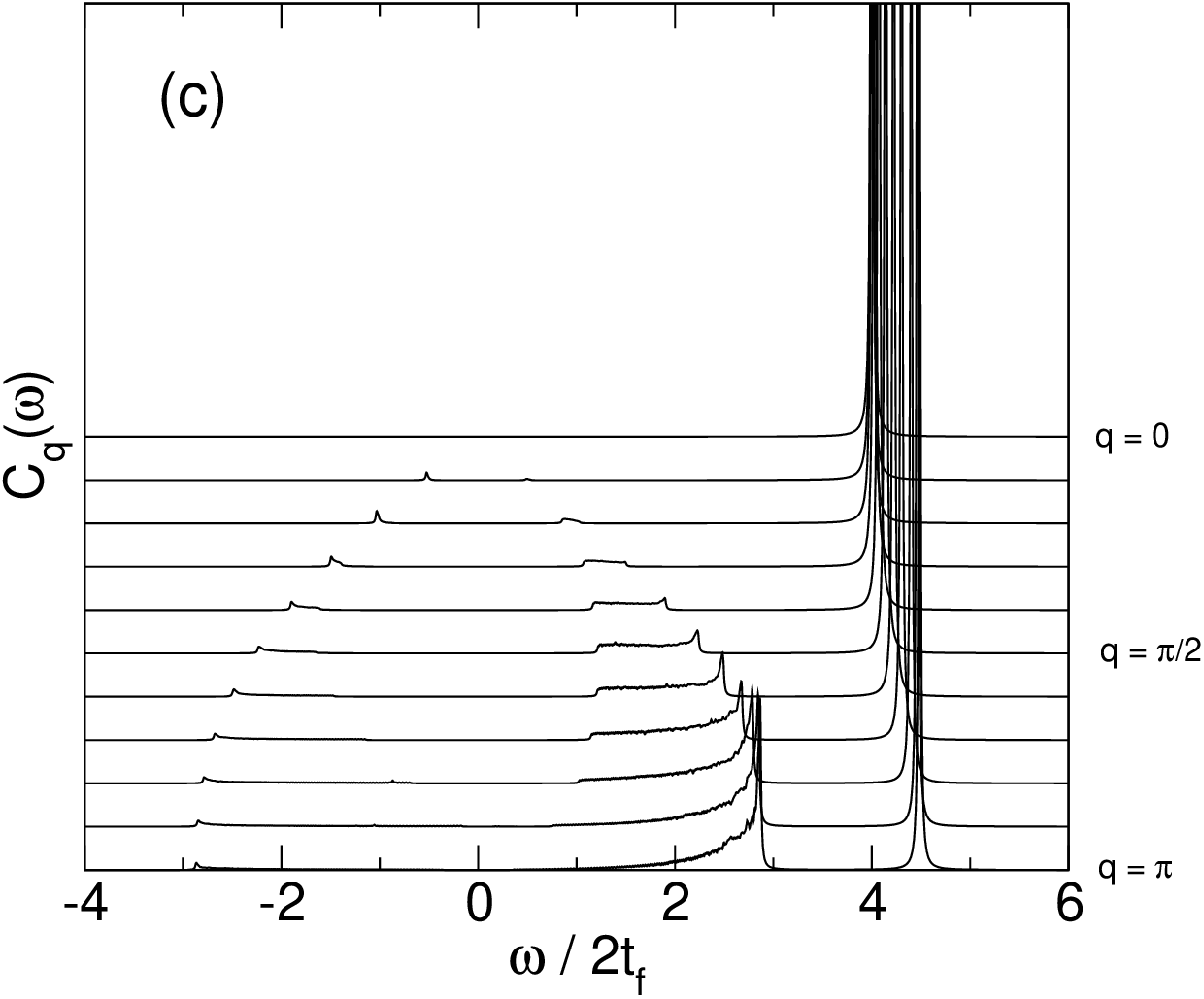}\hspace*{0.2cm}
\includegraphics[width=0.42\linewidth]{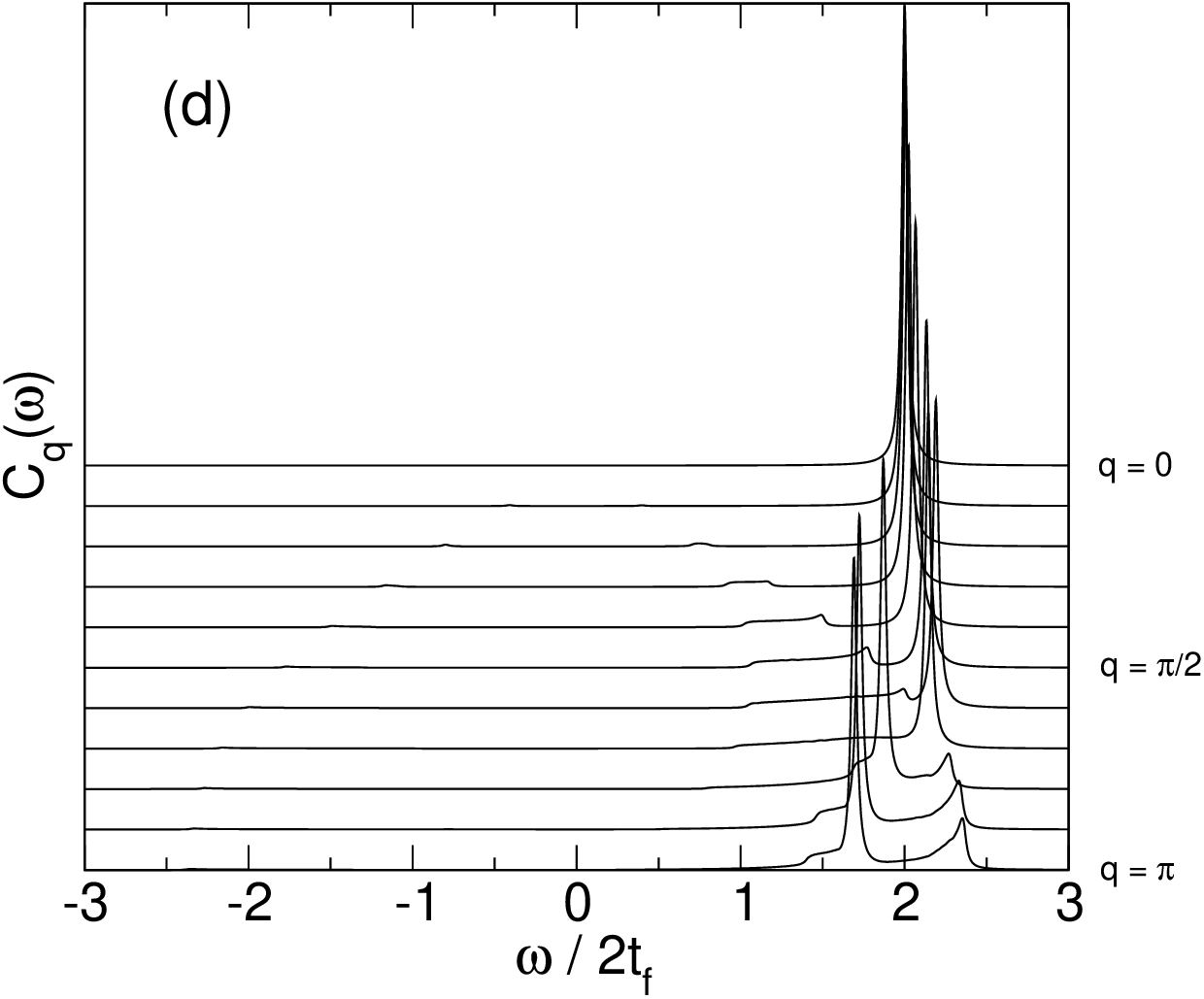}
%B_q_w4_tb2.eps %B_q_w2_tb1.eps
\caption{Boson spectral function $C_q(\omega)$ of the half-filled
fermion-boson transport model~\eqref{hem} with: $\Omega =10$, 
$\Lambda=1/4$ [panel (a)]; $\Omega =1/150$, 
$\Lambda=1/90$ [panel (b)]; $\Omega =4$,  
$\Lambda=1$ [panel (c)];  $\Omega =4$,  
$\Lambda=2$ [panel (d)]. The frequency $\omega$ is 
given in units of $2t_f$.}
\label{fig7}
\end{figure*}

\begin{figure}[hbt]
\includegraphics[width=0.9\linewidth]{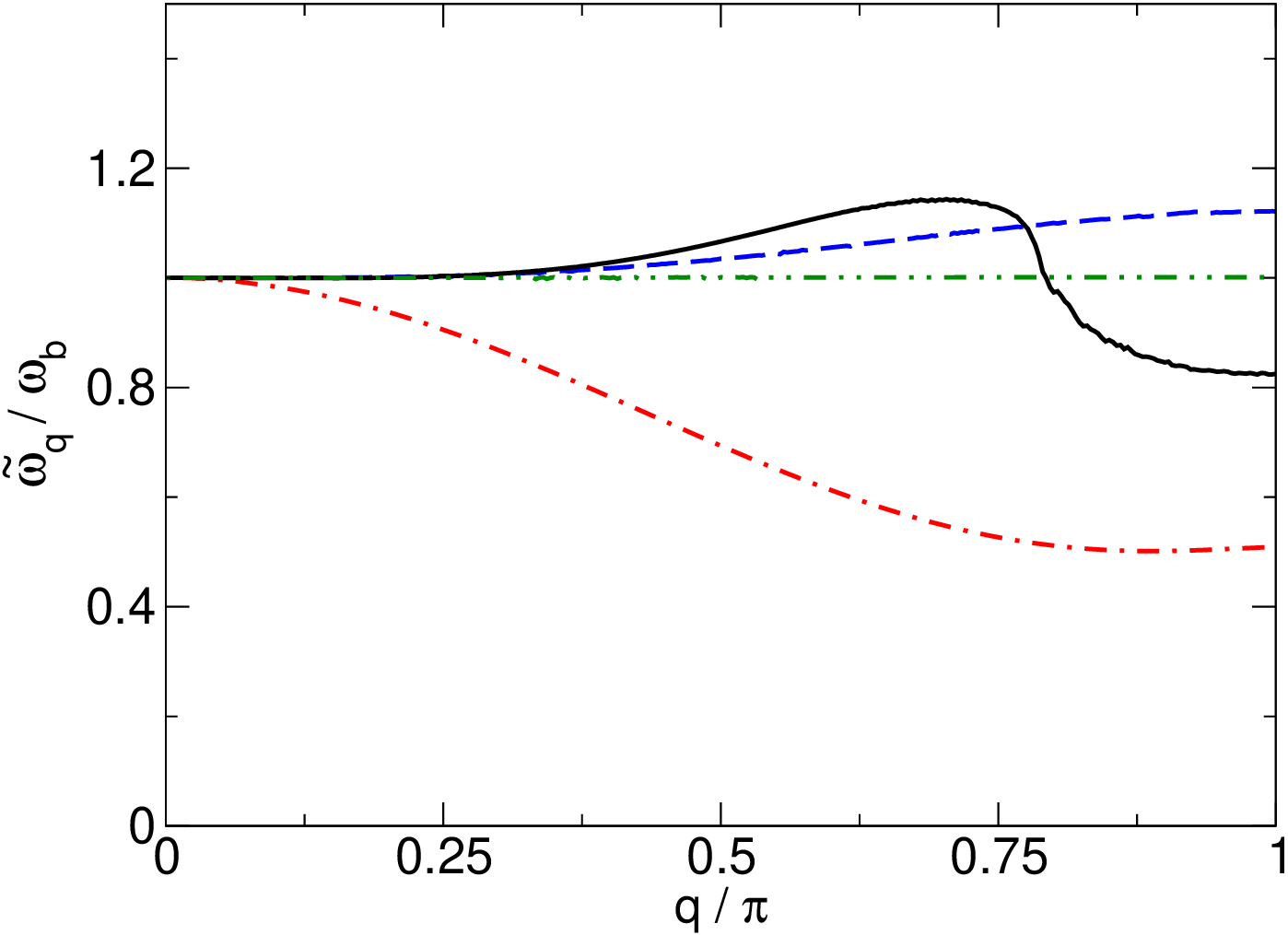}
%ren_e_w_tb1.eps
\caption{(Color online) Renormalized boson frequencies 
for the parameters used in Fig.~\ref{fig7}:
$\Omega =10$, $\Lambda=1/4$ (green double-dot-dashed line;
$\Omega =1/150$, $\Lambda=1/90$ (red dot-dashed line);
$\Omega =4$,  $\Lambda=1$ (blue dashed line),
$\Omega =4$,  $\Lambda=2$ (black solid line).}
\label{fig8}
\end{figure}

Figure~\ref{fig7} shows $C_q(\omega)$ in the metallic regime, for different
$\Omega$ and  $\Lambda$  parameters. For very large $\Omega=10$ [panel (a)], 
the boson energy is hardly renormalized by the coupling to the 
fermions. Accordingly we observe a strong signal at the 
bare boson frequency $\Omega/2t_f =100$ (first term in Eq.~\eqref{G5};
the second term in Eq.~\eqref{G5} will not contribute because there are 
no states available with $\omega=-\Omega/2t_f$). The third term in 
Eq.~\eqref{G5} detects particle-hole excitations and leads to the two
incoherent absorption bands in $C_q(\omega)$ running 
from $q=0\to\pi$ with energies between $\omega=0$ 
and $\omega\simeq \pm 10 t_f = \pm t_b/2$. At small $\Omega=0.00\bar{6}$, see  
panel (b), the (one-) boson excitation is located within the fermionic
band. As a result we find a strong renormalization of the bare
boson frequency (see also $\tilde{\omega}_q$ in Fig.~\ref{fig8}),
leading to the dispersive signal in the range $\omega/2t_f\simeq 
0.5\times 10^{-3}\ldots 10^{-3}$. We note that for $\Lambda =0.0\bar{1}$
used in panel~(b) the fermion-boson coupling is small in comparison with
the free fermion bandwidth (we have $g_k/\varepsilon_k=t_b/t_f=0.3$
in the model~\eqref{5}), hence the effect of multi-boson 
absorption processes is negligible. The lower two panels of 
Fig.~\ref{fig7} demonstrate how $\Lambda$ affects
the boson absorption at fixed $\Omega$. In panel (c),
for $\Lambda=1$ and $\Omega=4$, the boson frequency  is larger by 
a factor of two than the `free' fermion bandwidth ($4t_f=2$), 
whereas they have the same size for the  $C_q(\omega)$ spectrum with 
$\Lambda =2$ shown in panel (d). Quite differently, in the former case, 
the bare boson mode hardens, while it softens near $Q=\pi$ in the latter 
case, where the fermion and boson degrees of freedom are strongly mixed.
This becomes even more visible by comparing the corresponding
(dashed and solid) $\tilde{\omega}_q$ curves in Fig.~\ref{fig8}.  
  
\section{Conclusions}
To summarize, we adapted the projective renormalization method  
to the investigation of a novel two-channel fermion-boson model,
describing charge transport within a background medium. 
By large-scale numerical DMRG studies this model has been 
proven to show a metal-insulator quantum phase transition
for the one-dimensional half-filled band case. The transition is 
triggered by strong correlations evolving in the background and 
typifies as a Luttinger-liquid charge-density-wave crossover. 

Our analytical approach captures this TLL-CDW transition
for the infinite system, without truncating the bosonic Hilbert
space as in purely numerical investigation schemes. Therefore, 
the PRM is particularly well suited to analyze the bosonic 
degrees of freedom when passing the metal-insulator phase boundary. 

In the course of the renormalization procedure of the fermion-boson
Hamiltonian we end up with a model of noninteracting---but 
dressed---electrons and bosons. In this way, 
the renormalization of the fermion band dispersion  and 
of the boson frequency is obtained, and various
ground-state expectation values were calculated. 
Moreover, we derived analytical expressions for the 
single-particle (inverse) photoemission spectra and for the  
boson spectral function, which allows us to pinpoint the 
most important absorption and emission processes during particle 
transport, and throws some light on the nature of the 
TLL-CDW transition.

In particular, we show that the insulating CDW phase, realized for large
boson frequencies $\Omega$ and small boson relaxation parameter
$\Lambda$, is characterized by a gapful, mass 
asymmetric band structure. Thereby the lower
occupied band is almost flat by reason that transport is only possible 
through a vacuum restoring six-step hopping process. The CDW phase is a 
few-boson state. By contrast, the metallic (TLL) phase is a many-boson 
state, especially for small $\Omega$, where the background heavily 
fluctuates. Note that this regime is not that easy accessible by
numerical approaches. Here we observe a well-defined electron band 
in the vicinity of $k_F$ only. Due to many intersecting boson branches 
the (inverse) photoemission spectra are `overdamped' near the Brillouin 
zone boundaries.  

The boson spectral function and renormalized boson dispersion 
clearly indicate that the TLL-CDW transition is not accompanied
by a softening of the zone-boundary boson mode. Rather a harding of
the $Q=\pi$ boson is observed.  This might be partially attributed
to the vanishing Fourier-transformed fermion-boson coupling term at 
wave-numbers $\pm\pi/2$, which denote the two Fermi points for the 
half-filled band case. The situation changes if we look for
a metal-insulator transition for other commensurate band-filling 
factors, e.g., at quarter filling. Whether the system there undergoes 
a soft-mode transition for small $\Omega$ is an interesting open question
that deserves future efforts. In this connection other kinds of
charge-ordered states should be also considered.

%------------------------------------------------------------------------------
\section*{Acknowledgments}
The authors would like to thank A. Alvermann, D.~M. Edwards, S. Ejima, and 
A.~H\"ubsch for many stimulating and enlightening discussions. 
This work was supported by the DFG through the research 
program SFB 652.

%%%%%%%%%%%%%%%%%%%%%%%%%%%%%%%%%%%%%%%%%%%%%%%%%%%%%%%%%%%%%%%%%%%%%%%%%%%%%%%

%\bibliographystyle{apsrev}
%\bibliography{./ref} 

\begin{thebibliography}{31}
\expandafter\ifx\csname natexlab\endcsname\relax\def\natexlab#1{#1}\fi
\expandafter\ifx\csname bibnamefont\endcsname\relax
  \def\bibnamefont#1{#1}\fi
\expandafter\ifx\csname bibfnamefont\endcsname\relax
  \def\bibfnamefont#1{#1}\fi
\expandafter\ifx\csname citenamefont\endcsname\relax
  \def\citenamefont#1{#1}\fi
\expandafter\ifx\csname url\endcsname\relax
  \def\url#1{\texttt{#1}}\fi
\expandafter\ifx\csname urlprefix\endcsname\relax\def\urlprefix{URL }\fi
\providecommand{\bibinfo}[2]{#2}
\providecommand{\eprint}[2][]{\url{#2}}

\bibitem[{\citenamefont{Gr{\"u}ner}(1994)}]{Gr94}
\bibinfo{author}{\bibfnamefont{G.}~\bibnamefont{Gr{\"u}ner}},
  \emph{\bibinfo{title}{Density Waves in Solids}} (\bibinfo{publisher}{Addison
  Wesley, Reading, MA}, \bibinfo{year}{1994}).

\bibitem[{\citenamefont{Bishop and Swanson}(1993)}]{BS93}
\bibinfo{author}{\bibfnamefont{A.~R.} \bibnamefont{Bishop}} \bibnamefont{and}
  \bibinfo{author}{\bibfnamefont{B.~I.} \bibnamefont{Swanson}},
  \bibinfo{journal}{Los Alamos Sciences} \textbf{\bibinfo{volume}{21}},
  \bibinfo{pages}{133} (\bibinfo{year}{1993}).

\bibitem[{\citenamefont{Su et~al.}(1979)\citenamefont{Su, Schrieffer, and
  Heeger}}]{SSH79}
\bibinfo{author}{\bibfnamefont{W.~P.} \bibnamefont{Su}},
  \bibinfo{author}{\bibfnamefont{J.~R.} \bibnamefont{Schrieffer}},
  \bibnamefont{and} \bibinfo{author}{\bibfnamefont{A.~J.}
  \bibnamefont{Heeger}}, \bibinfo{journal}{Phys. Rev. Lett.}
  \textbf{\bibinfo{volume}{42}}, \bibinfo{pages}{1698} (\bibinfo{year}{1979}).

\bibitem[{\citenamefont{Holstein}(1959)}]{Ho59a}
\bibinfo{author}{\bibfnamefont{T.}~\bibnamefont{Holstein}},
  \bibinfo{journal}{Ann. Phys. (N.Y.)} \textbf{\bibinfo{volume}{8}},
  \bibinfo{pages}{325} (\bibinfo{year}{1959}).

\bibitem[{\citenamefont{Bursill et~al.}(1998)\citenamefont{Bursill, McKenzie,
  and Hamer}}]{BMH98}
\bibinfo{author}{\bibfnamefont{R.~J.} \bibnamefont{Bursill}},
  \bibinfo{author}{\bibfnamefont{R.~H.} \bibnamefont{McKenzie}},
  \bibnamefont{and} \bibinfo{author}{\bibfnamefont{C.~J.} \bibnamefont{Hamer}},
  \bibinfo{journal}{Phys. Rev. Lett.} \textbf{\bibinfo{volume}{80}},
  \bibinfo{pages}{5607} (\bibinfo{year}{1998}).

\bibitem[{\citenamefont{Hohenadler et~al.}(2006)\citenamefont{Hohenadler,
  Wellein, Bishop, Alvermann, and Fehske}}]{HWBAF06}
\bibinfo{author}{\bibfnamefont{M.}~\bibnamefont{Hohenadler}},
  \bibinfo{author}{\bibfnamefont{G.}~\bibnamefont{Wellein}},
  \bibinfo{author}{\bibfnamefont{A.~R.} \bibnamefont{Bishop}},
  \bibinfo{author}{\bibfnamefont{A.}~\bibnamefont{Alvermann}},
  \bibnamefont{and} \bibinfo{author}{\bibfnamefont{H.}~\bibnamefont{Fehske}},
  \bibinfo{journal}{Phys. Rev. B} \textbf{\bibinfo{volume}{73}},
  \bibinfo{pages}{245120} (\bibinfo{year}{2006}).

\bibitem[{\citenamefont{Peierls}(1955)}]{Pe55}
\bibinfo{author}{\bibfnamefont{R.}~\bibnamefont{Peierls}},
  \emph{\bibinfo{title}{Quantum theory of solids}} (\bibinfo{publisher}{Oxford
  University Press}, \bibinfo{address}{Oxford}, \bibinfo{year}{1955}).

\bibitem[{\citenamefont{Hubbard}(1963)}]{Hu63}
\bibinfo{author}{\bibfnamefont{J.}~\bibnamefont{Hubbard}},
  \bibinfo{journal}{Proc. Roy. Soc. London, Ser. A}
  \textbf{\bibinfo{volume}{276}}, \bibinfo{pages}{238} (\bibinfo{year}{1963}).

\bibitem[{\citenamefont{Mutou et~al.}(1998)\citenamefont{Mutou, Shibata, and
  Ueda}}]{MSU98}
\bibinfo{author}{\bibfnamefont{T.}~\bibnamefont{Mutou}},
  \bibinfo{author}{\bibfnamefont{N.}~\bibnamefont{Shibata}}, \bibnamefont{and}
  \bibinfo{author}{\bibfnamefont{K.}~\bibnamefont{Ueda}},
  \bibinfo{journal}{Phys. Rev. B} \textbf{\bibinfo{volume}{57}},
  \bibinfo{pages}{13702} (\bibinfo{year}{1998}).

\bibitem[{\citenamefont{Takada and Chatterjee}(2003)}]{TC03}
\bibinfo{author}{\bibfnamefont{Y.}~\bibnamefont{Takada}} \bibnamefont{and}
  \bibinfo{author}{\bibfnamefont{A.}~\bibnamefont{Chatterjee}},
  \bibinfo{journal}{Phys. Rev. B} \textbf{\bibinfo{volume}{67}},
  \bibinfo{pages}{081102(R)} (\bibinfo{year}{2003});
%\bibitem[{\citenamefont{Fehske et~al.}(2001)\citenamefont{Fehske, Kinateder,
%  Wellein, and Bishop}}]{FKWB01}
\bibinfo{author}{\bibfnamefont{H.}~\bibnamefont{Fehske}},
  \bibinfo{author}{\bibfnamefont{M.}~\bibnamefont{Kinateder}},
  \bibinfo{author}{\bibfnamefont{G.}~\bibnamefont{Wellein}}, \bibnamefont{and}
  \bibinfo{author}{\bibfnamefont{A.~R.} \bibnamefont{Bishop}},
  \bibinfo{journal}{Phys. Rev. B} \textbf{\bibinfo{volume}{63}},
  \bibinfo{pages}{245121} (\bibinfo{year}{2001});
%\bibitem[{\citenamefont{Fehske et~al.}(2004)\citenamefont{Fehske, Wellein,
%  Hager, Wei{\ss}e, and Bishop}}]{FWHWB04}
\bibinfo{author}{\bibfnamefont{H.}~\bibnamefont{Fehske}},
  \bibinfo{author}{\bibfnamefont{G.}~\bibnamefont{Wellein}},
  \bibinfo{author}{\bibfnamefont{G.}~\bibnamefont{Hager}},
  \bibinfo{author}{\bibfnamefont{A.}~\bibnamefont{Wei{\ss}e}},
  \bibnamefont{and} \bibinfo{author}{\bibfnamefont{A.~R.}
  \bibnamefont{Bishop}}, \bibinfo{journal}{Phys. Rev. B}
  \textbf{\bibinfo{volume}{69}}, \bibinfo{pages}{165115}
  (\bibinfo{year}{2004});
%\bibitem[{\citenamefont{Clay and Hardikar}(2005)}]{CH05}
\bibinfo{author}{\bibfnamefont{R.~T.} \bibnamefont{Clay}} \bibnamefont{and}
  \bibinfo{author}{\bibfnamefont{R.~P.} \bibnamefont{Hardikar}},
  \bibinfo{journal}{Phys. Rev. Lett.} \textbf{\bibinfo{volume}{95}},
  \bibinfo{pages}{096401} (\bibinfo{year}{2005});
%\bibitem[{\citenamefont{Bissola and Parola}(2006)}]{BP06}
\bibinfo{author}{\bibfnamefont{S.}~\bibnamefont{Bissola}} \bibnamefont{and}
  \bibinfo{author}{\bibfnamefont{A.}~\bibnamefont{Parola}},
  \bibinfo{journal}{Phys. Rev. B} \textbf{\bibinfo{volume}{73}},
  \bibinfo{pages}{195108} (\bibinfo{year}{2006}).

\bibitem[{\citenamefont{Wohlfeld et~al.}(2009)\citenamefont{Wohlfeld, Ole\'{s},
  and Horsch}}]{WOH09}
\bibinfo{author}{\bibfnamefont{K.}~\bibnamefont{Wohlfeld}},
  \bibinfo{author}{\bibfnamefont{A.~M.} \bibnamefont{Ole\'{s}}},
  \bibnamefont{and} \bibinfo{author}{\bibfnamefont{P.}~\bibnamefont{Horsch}},
  \bibinfo{journal}{Phys. Rev. B} \textbf{\bibinfo{volume}{79}},
  \bibinfo{pages}{224433} (\bibinfo{year}{2009}).

\bibitem[{\citenamefont{Berciu}(2009)}]{Be09}
\bibinfo{author}{\bibfnamefont{M.}~\bibnamefont{Berciu}},
  \bibinfo{journal}{Physics} \textbf{\bibinfo{volume}{2}}, \bibinfo{pages}{55}
  (\bibinfo{year}{2009}).

\bibitem[{\citenamefont{Komineas et~al.}(2007)\citenamefont{Komineas, Kalosaka,
  and Bishop}}]{KKB02}
\bibinfo{author}{\bibfnamefont{S.}~\bibnamefont{Komineas}},
  \bibinfo{author}{\bibfnamefont{G.}~\bibnamefont{Kalosaka}}, \bibnamefont{and}
  \bibinfo{author}{\bibfnamefont{A.~R.} \bibnamefont{Bishop}},
  \bibinfo{journal}{Phys. Rev. E} \textbf{\bibinfo{volume}{65}},
  \bibinfo{pages}{061905} (\bibinfo{year}{2007}).

\bibitem[{\citenamefont{Edwards}(2006)}]{Ed06}
\bibinfo{author}{\bibfnamefont{D.~M.} \bibnamefont{Edwards}},
  \bibinfo{journal}{Physica B} \textbf{\bibinfo{volume}{378-380}},
  \bibinfo{pages}{133} (\bibinfo{year}{2006}).

\bibitem[{\citenamefont{Alvermann et~al.}(2007)\citenamefont{Alvermann,
  Edwards, and Fehske}}]{AEF07}
\bibinfo{author}{\bibfnamefont{A.}~\bibnamefont{Alvermann}},
  \bibinfo{author}{\bibfnamefont{D.~M.} \bibnamefont{Edwards}},
  \bibnamefont{and} \bibinfo{author}{\bibfnamefont{H.}~\bibnamefont{Fehske}},
  \bibinfo{journal}{Phys. Rev. Lett.} \textbf{\bibinfo{volume}{98}},
  \bibinfo{pages}{056602} (\bibinfo{year}{2007}).

\bibitem[{\citenamefont{Wellein et~al.}(2008)\citenamefont{Wellein, Fehske,
  Alvermann, and Edwards}}]{WFAE08}
\bibinfo{author}{\bibfnamefont{G.}~\bibnamefont{Wellein}},
  \bibinfo{author}{\bibfnamefont{H.}~\bibnamefont{Fehske}},
  \bibinfo{author}{\bibfnamefont{A.}~\bibnamefont{Alvermann}},
  \bibnamefont{and} \bibinfo{author}{\bibfnamefont{D.~M.}
  \bibnamefont{Edwards}}, \bibinfo{journal}{Phys. Rev. Lett.}
  \textbf{\bibinfo{volume}{101}}, \bibinfo{pages}{136402}
  (\bibinfo{year}{2008}).

\bibitem[{\citenamefont{Ejima et~al.}(2009)\citenamefont{Ejima, Hager, and
  Fehske}}]{EHF09}
\bibinfo{author}{\bibfnamefont{S.}~\bibnamefont{Ejima}},
  \bibinfo{author}{\bibfnamefont{G.}~\bibnamefont{Hager}}, \bibnamefont{and}
  \bibinfo{author}{\bibfnamefont{H.}~\bibnamefont{Fehske}},
  \bibinfo{journal}{Phys. Rev. Lett.} \textbf{\bibinfo{volume}{102}},
  \bibinfo{pages}{106404} (\bibinfo{year}{2009}).

\bibitem[{\citenamefont{Ejima and Fehske}(2009)}]{EF09b}
\bibinfo{author}{\bibfnamefont{S.}~\bibnamefont{Ejima}} \bibnamefont{and}
  \bibinfo{author}{\bibfnamefont{H.}~\bibnamefont{Fehske}},
  \bibinfo{journal}{Phys. Rev. B} \textbf{\bibinfo{volume}{80}},
  \bibinfo{pages}{155101} (\bibinfo{year}{2009}).

\bibitem[{\citenamefont{H\"ubsch et~al.}(2008)\citenamefont{H\"ubsch, Sykora,
  and Becker}}]{HSB08}
\bibinfo{author}{\bibfnamefont{A.}~\bibnamefont{H\"ubsch}},
  \bibinfo{author}{\bibfnamefont{S.}~\bibnamefont{Sykora}}, \bibnamefont{and}
  \bibinfo{author}{\bibfnamefont{K.~W.} \bibnamefont{Becker}}
  (\bibinfo{year}{2008}), \bibinfo{note}{preprint},
  \urlprefix\url{arXiv:0809.3360};
%\bibitem[{\citenamefont{Becker et~al.}(2002)\citenamefont{Becker, H\"ubsch, and
%  Sommer}}]{BHS02}
\bibinfo{author}{\bibfnamefont{K.~W.} \bibnamefont{Becker}},
  \bibinfo{author}{\bibfnamefont{A.}~\bibnamefont{H\"ubsch}}, \bibnamefont{and}
  \bibinfo{author}{\bibfnamefont{T.}~\bibnamefont{Sommer}},
  \bibinfo{journal}{Phys. Rev. B} \textbf{\bibinfo{volume}{66}},
  \bibinfo{pages}{235115} (\bibinfo{year}{2002}).

\bibitem[{\citenamefont{Wegner}(1994)}]{We94}
\bibinfo{author}{\bibfnamefont{F.}~\bibnamefont{Wegner}},
  \bibinfo{journal}{Ann. Phys. (Leipzig)} \textbf{\bibinfo{volume}{3}},
  \bibinfo{pages}{77} (\bibinfo{year}{1994}).

\bibitem[{\citenamefont{H\"ubsch and Becker}(2005)}]{HB05}
\bibinfo{author}{\bibfnamefont{A.}~\bibnamefont{H\"ubsch}} \bibnamefont{and}
  \bibinfo{author}{\bibfnamefont{K.~W.} \bibnamefont{Becker}},
  \bibinfo{journal}{Phys. Rev. B} \textbf{\bibinfo{volume}{71}},
  \bibinfo{pages}{155116} (\bibinfo{year}{2005}).

\bibitem[{\citenamefont{Sykora et~al.}(2005)\citenamefont{Sykora, H{\"u}bsch,
  Becker, Wellein, and Fehske}}]{SHBWF05}
\bibinfo{author}{\bibfnamefont{S.}~\bibnamefont{Sykora}},
  \bibinfo{author}{\bibfnamefont{A.}~\bibnamefont{H{\"u}bsch}},
  \bibinfo{author}{\bibfnamefont{K.~W.} \bibnamefont{Becker}},
  \bibinfo{author}{\bibfnamefont{G.}~\bibnamefont{Wellein}}, \bibnamefont{and}
  \bibinfo{author}{\bibfnamefont{H.}~\bibnamefont{Fehske}},
  \bibinfo{journal}{Phys. Rev. B} \textbf{\bibinfo{volume}{71}},
  \bibinfo{pages}{045112} (\bibinfo{year}{2005}).



\bibitem[{\citenamefont{Becker et~al.}(2007{\natexlab{b}})\citenamefont{Becker,
  Sykora, and Zlatic}}]{BSZ07}
\bibinfo{author}{\bibfnamefont{K.~W.} \bibnamefont{Becker}},
\bibinfo{author}{\bibfnamefont{S.}~\bibnamefont{Sykora}},
  \bibnamefont{and}
  \bibinfo{author}{\bibfnamefont{V.} \bibnamefont{Zlatic}},
  \bibinfo{journal}{Phys. Rev. B} \textbf{\bibinfo{volume}{75}},
  \bibinfo{pages}{075101} (\bibinfo{year}{2007}{\natexlab{b}}).

\bibitem{SEprivate}
S. Ejima, private communication.

\bibitem[{\citenamefont{Sykora et~al.}(2009)\citenamefont{Sykora, H\"ubsch, and
  Becker}}]{SHB09}
\bibinfo{author}{\bibfnamefont{S.}~\bibnamefont{Sykora}},
  \bibinfo{author}{\bibfnamefont{A.}~\bibnamefont{H\"ubsch}}, \bibnamefont{and}
  \bibinfo{author}{\bibfnamefont{K.~W.} \bibnamefont{Becker}},
  \bibinfo{journal}{Europhys. Lett.} \textbf{\bibinfo{volume}{85}},
  \bibinfo{pages}{57003} (\bibinfo{year}{2009}).
%\bibitem{SSykoraprivate}
%S. Sykora, private communication

\bibitem[{\citenamefont{Sykora et~al.}(2006)\citenamefont{Sykora,
   H\"ubsch, and Becker}}]{SHB06}
 \bibinfo{author}{\bibfnamefont{S.}~\bibnamefont{Sykora}},
   \bibinfo{author}{\bibfnamefont{A.}~\bibnamefont{H\"ubsch}}, \bibnamefont{and}
   \bibinfo{author}{\bibfnamefont{K.~W.} \bibnamefont{Becker}},
   \bibinfo{journal}{Europhys. Lett.} \textbf{\bibinfo{volume}{76}},
   \bibinfo{pages}{644} (\bibinfo{year}{2006}).

\end{thebibliography}

\end{document}